%% file: ESOs_Star_Tail.tex
\documentclass[fleqn,usenatbib]{mnras}

\usepackage[T1]{fontenc}
\usepackage{ae,aecompl}

\usepackage{graphicx}	 
\usepackage{epstopdf}   
\usepackage{amsmath}	 
\usepackage{amssymb}	 
\usepackage[shortcuts]{extdash}    
\usepackage{threeparttable}
\usepackage[dvipsnames]{xcolor}
\usepackage[acronym]{glossaries}   

\usepackage{newtxtext,newtxmath}

\graphicspath{{graphics/}
              {images/}}
\DeclareGraphicsExtensions{.pdf,.png}
\newcommand{\eso}{ESO~137\=/001}       
\newcommand{\hst}{\emph{HST}}          
\newcommand{\chandra}{\emph{Chandra}}  

\glsdisablehyper
\newacronym{agn}{AGN}{active galactic nucleus}
\newacronym{alma}{\emph{ALMA}}{the Atacama Large Millimeter Array}
\newacronym{atca}{\emph{ATCA}}{Australia Telescope Compact Array}
\newacronym{acs}{ACS}{Advanced Camera System}
\newacronym{cr}{CR}{cosmic ray}
\newacronym{ew}{EW}{equivalent width}
\newacronym[description={color-color \gls{sf}}]{esf}{cc\=/SF}{color-color SF}
\newacronym[description={color-color \gls{sfr}}]{esfr}{cc\=/SFR}{color-color SFR}
\newacronym[longplural={fields of view}]{fov}{FOV}{field of view}
\newacronym{fwhm}{FWHM}{full width at half maximum}
\newacronym{gasp}{GASP}{gas stripping phenomena}
\newacronym{kde}{KDE}{kernel density estimation}
\newacronym{hst}{\hst{}}{\emph{Hubble Space Telescope}}
\newacronym{imf}{IMF}{initial mass function}
\newacronym{icm}{ICM}{intracluster medium}
\newacronym{ism}{ISM}{interstellar medium}
\newacronym{ivm}{IVM}{inverse variance map}
\newacronym{mad}{MAD}{median absolute deviation}
\newacronym{mcmc}{MCMC}{Markov chain Monte Carlo}
\newacronym{muse}{\emph{MUSE}}{Multi Unit Spectroscopic Explorer}
\newacronym{psf}{PSF}{point spread function}
\newacronym{rms}{RMS}{root mean square}
\newacronym{rps}{RPS}{ram pressure stripping}
\newacronym{soar}{\emph{SOAR}}{Southern Observatory for Astrophysical Research}
\newacronym{sf}{SF}{star formation}
\newacronym{sfr}{SFR}{star formation rate}
\newacronym{ssp}{SSP}{simple stellar population}
\newacronym{ulx}{ULX}{ultraluminous X-ray source}
\newacronym{wfc3}{WFC3}{Wide Field Camera~3}

\defcitealias{sun2006}{S06}
\defcitealias{sun2007HA}{S07}
\defcitealias{sun2010}{S10}
\defcitealias{fossati2016}{F16} 

\newcommand{\hi}{H\,{\sc i}}
\newcommand{\hii}{H\,{\sc ii}}

\title[Star-forming trails behind \eso{}]{\hst{} viewing of spectacular star-forming trails behind \eso{}}

\author[W. Waldron et al.]
{William Waldron$^{1,2}$\thanks{E-mail: wwaldron@harding.edu},
Ming Sun$^{1}$\thanks{E-mail: ms0071@uah.edu},
Rongxin Luo$^{1}$\thanks{E-mail: rl0055@uah.edu},
Sunil Laudari$^{1}$,
Marios Chatzikos$^{3}$,
\newauthor
Suresh Sivanandam$^{4,5}$,
Jeffrey D. P. Kenney$^{6}$,
Pavel J\'{a}chym$^{7}$,
G. Mark Voit$^{8}$,
Megan Donahue$^{8}$,
\newauthor
Matteo Fossati$^{9,10}$
\\
$^{1}$Department of Physics, University of Alabama in Huntsville, 301 Sparkman Dr NW, Huntsville, AL 35899, USA\\
$^{2}$Engineering \& Physics Department, Harding University, 915 E. Market Ave, Searcy, AR 72143, USA\\
$^{3}$Department of Physics \& Astronomy, University of Kentucky, Lexington, KY 40506, USA\\
$^{4}$David A. Dunlap Department of Astronomy and Astrophysics, University of Toronto, 50 St. George St, Toronto, ON, Canada\\
$^{5}$Dunlap Institute for Astronomy and Astrophysics, University of Toronto, 50 St. George St, Toronto, ON, Canada\\
$^{6}$Yale University Astronomy Department, P.O. Box 208101, New Haven, CT 06520-8101 USA\\
$^{7}$Astronomical Institute of the Czech Academy of Sciences, Bo\v{c}n\'{i} II 1401, 141 00, Prague, Czech Republic\\
$^{8}$Department of Physics and Astronomy, Michigan State University, East Lansing, MI 48824, USA\\
$^{9}$Dipartimento di Fisica G. Occhialini, Universit\`{a} degli Studi di Milano Bicocca, Piazza della Scienza 3, I-20126 Milano, Italy \\
$^{10}$INAF-Osservatorio Astronomico di Brera, via Brera 28, I-20121 Milano, Italy \\
}

\date{Accepted XXX. Received YYY; in original form ZZZ}

\pubyear{2021}

\begin{document}
\label{firstpage}
\pagerange{\pageref{firstpage}--\pageref{lastpage}}
\maketitle

\begin{abstract}
We present the results from the \acrshort{hst} \acrshort{wfc3} and \acrshort{acs} data on an archetypal galaxy undergoing \acrfull{rps}, \eso{}, in the nearby cluster Abell~3627. \eso{} is known to host a prominent stripped tail detected in many bands from X-rays, H$\alpha$ to CO. The \acrshort{hst} data reveal significant features indicative of \acrshort{rps} such as asymmetric dust distribution and surface brightness as well as many blue young star complexes in the tail. We study the correlation between the blue young star complexes from \acrshort{hst}, \ion{H}{ii} regions from H$\alpha$ (\acrshort{muse}) and dense molecular clouds from CO (\acrshort{alma}). The correlation between the \acrshort{hst} blue star clusters and the \ion{H}{ii} regions is very good, while their correlation with the dense CO clumps are typically not good, presumably due in part to evolutionary effects.
In comparison to the Starburst99+Cloudy model, many blue regions are found to be young ($<$~10\,Myr) and the total \acrfull{sf} rate in the tail is 0.3 - 0.6\,M$_{\sun{}}$/yr for sources measured with ages less than~100\,Myr, about 40\% of the \acrshort{sf} rate in the galaxy.
We trace SF over at least 100 Myr and give a full picture of the recent SF history in the tail.
We also demonstrate the importance of including nebular emissions and a nebular to stellar extinction correction factor when comparing the model to the broadband data. 
Our work on \eso{} demonstrates the importance of HST data for constraining the SF history in stripped tails.
\end{abstract}


\input{sections/s1-intro/intro}

\input{sections/s2-hst/hst}

\input{sections/s3-eso/eso}

\input{sections/s4-youngclusters/h2_regs}

\input{sections/s6-discuss/discuss}

\input{sections/s7-conc/conclude}

\input{sections/acknowledge}

\bibliographystyle{mnras}
\bibliography{bibliography}

\appendix

\input{sections/a1-phot/phot}

\bsp	
\label{lastpage}
\end{document}

%% file: sections/s1-intro/intro.tex
\section{Introduction}\label{sec:intro}

It has been long known that the environments where galaxies reside affect their properties and evolution \citep[e.g.,][]{dressler1980,Boselli2022}. 
Galaxies can be shaped by gravitational processes, e.g., mergers and tidal stripping.
Another important process that shapes galaxies, especially in galaxy clusters and groups, is \acrlong{rps} \citep{gunn1972,Boselli2022}. \Gls{rps} on galaxies comes from the drag force from the surrounding medium that is
proportional to its density and the square of the galaxy velocity.
\gls{rps} can impact galaxy evolution by first compressing the disk's \gls{ism} triggering
an initial star burst \citep[e.g.,][]{bekki2003}.
Once \gls{rps} sweeps through the galaxy, the cold \gls{ism} will be removed, eventually quenching star formation in the galaxy \citep[e.g.][]{quilis2000}.
\gls{rps} can have a significant impact on the properties of galaxies, e.g., disk truncation, formation of flocculent arms, transformation of dwarf galaxies and nuclear activity (see recent review by \citealt{Boselli2022}).
In recent years, evidence of \gls{rps} has been found in many clusters across different bands on different gas tracers from \hi, H$\alpha$, warm H$_{2}$, X-rays to CO (see the recent review by \citealt{Boselli2022}). It has been gradually established that the cold \gls{ism}, once stripped from galaxies, will mix with the surrounding hot \gls{icm}.
During the process, a fraction of the stripped \gls{ism} can cool to form stars in the tails,
while the bulk will
eventually be evaporated \citep[e.g.,][]{sun2021}. Nevertheless, many outstanding questions remain, e.g., the impact of RPS on galaxy properties, the \gls{sf} efficiency in tails, roles of magnetic field and turbulence, and the spatial and kinematic connection for gas of different phases. To address these questions, a multi-wavelength approach is required. Among all the multi-wavelength data, optical/UV broad-band data with \hst{} present the sharpest view of \gls{sf} in the stripped tails and allow detailed constraints on the ``sink term'' of the multi-phase stripped tails.

This work presents this type of study with the \gls{hst} data, on the archetypal \gls{rps} galaxy, \eso{}, also the one with the brightest diffuse H$\alpha$ tail and the richest amount of multi-wavelength supporting data \citep{sun2007HA,sun2010,siv2010,fumagalli2014,jachym2014,fossati2016,jachym2019,sun2021}.
It is also the only RPS galaxy in the \emph{JWST} GTO program with 14.5 hours of \emph{MIRI} observations (program 1178).
Here we study the impact of \gls{rps} on the galaxy.
We intend to quantify the \gls{sfr} in the tail and compare the \gls{hst} results with those from H$\alpha$ observations (as similarly done in e.g., \citealt{cramer2019} and \citealt{poggianti2019-g13}).
The \gls{sfr} history in the tail is also examined, as well as connection between young star clusters/complexes, \ion{H}{ii}, regions and CO clouds.

\input{sections/s1-intro/subs/eso}

%% file: sections/s1-intro/subs/eso.tex
\begin{figure*}
    \includegraphics[width=0.97\textwidth]{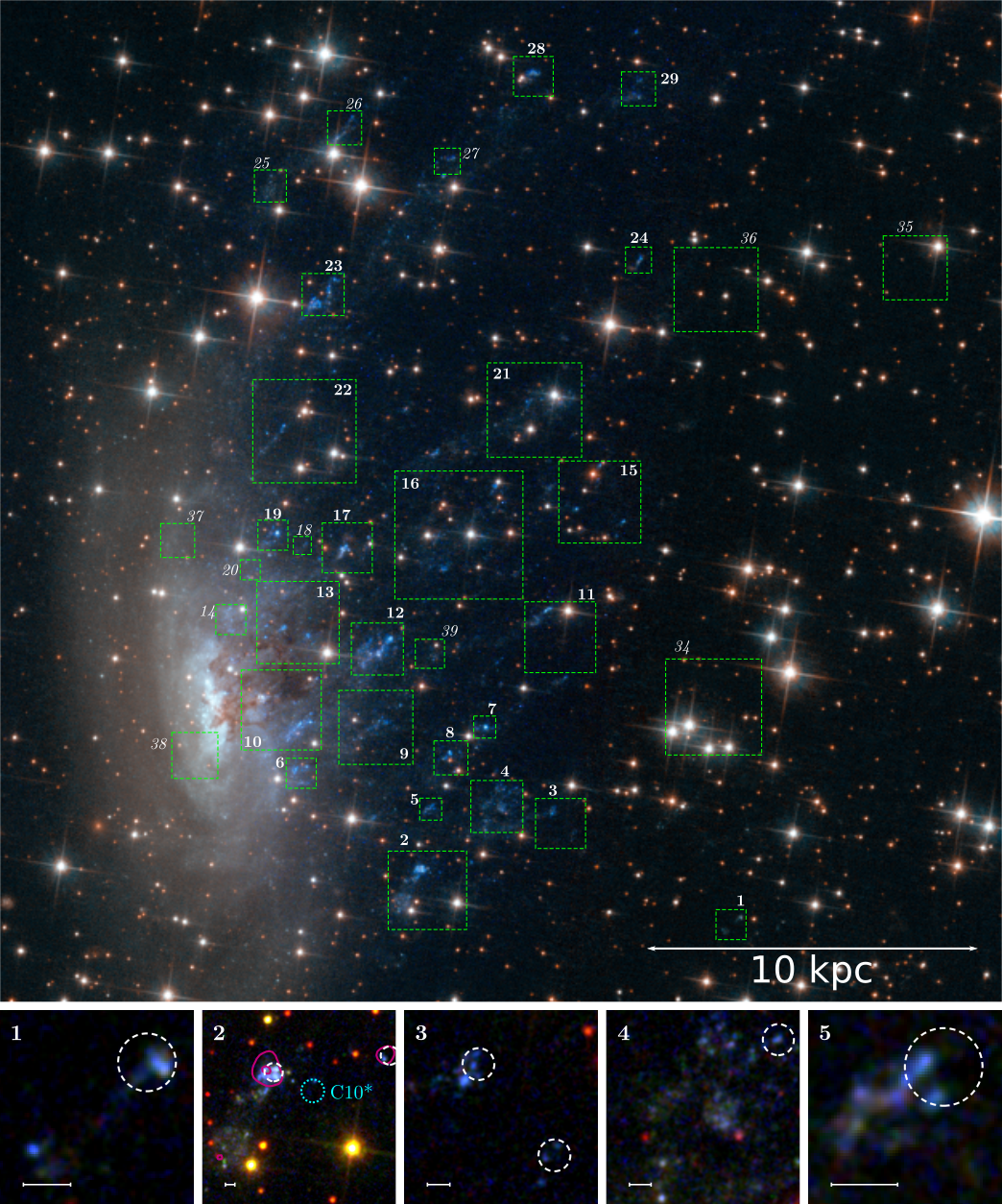}
    \vspace{-5pt}
    \caption{The RGB image (blue: F275W;
    green: F475W; red: F814W) of \eso{} with zoom-ins on~40 regions (Large Image Credit: STScI; zoom-ins are from this work). We include the following zoom-in regions: \ion{H}{II} regions defined in Section~\ref{sub:roi} shown by the white, dashed circles with a radius of 0\farcs{}40, X-ray point sources from \citet{sun2010} shown by the cyan, dotted circles with a radius of 0\farcs{}50, bright CO sources from \citet{jachym2019} shown by the magenta, solid contours (contour levels of~0.03,~0.06 and~0.12\,Jy\,km/s per beam), and blue star clusters defined from this work. The blue star clusters shown were selected with these criteria: F275W-F475W~<~2, F475W-F814W~<~1 and F475W~<~25.6. Some isolated blue sources are not included in zoom-ins but they are all studied in this work.
    The arrow on each cutout represents 0.25\,kpc. The bold numbers emphasize regions that align
    with an \ion{H}{II} source, and italicized numbers emphasize regions that do not.
    Regions~30,~31,~32,~33, and~40 lie outside the large STScI image and are shown for
    reference on Fig.~\ref{fig:srcPositions}. Region~40 is not covered by F275W so the blue source associated with the \hii{} region appears green.
    The correlation of \ion{H}{II}, blue \hst{} star clusters, CO clumps and X-ray sources is discussed in Section~\ref{sec:youngClusters}. 
    }
    \label{fig:wholeWithRegs}
\end{figure*}
\begin{figure*}
    \centering
    \vspace{-0.3cm}
    \includegraphics[width=0.97\textwidth]{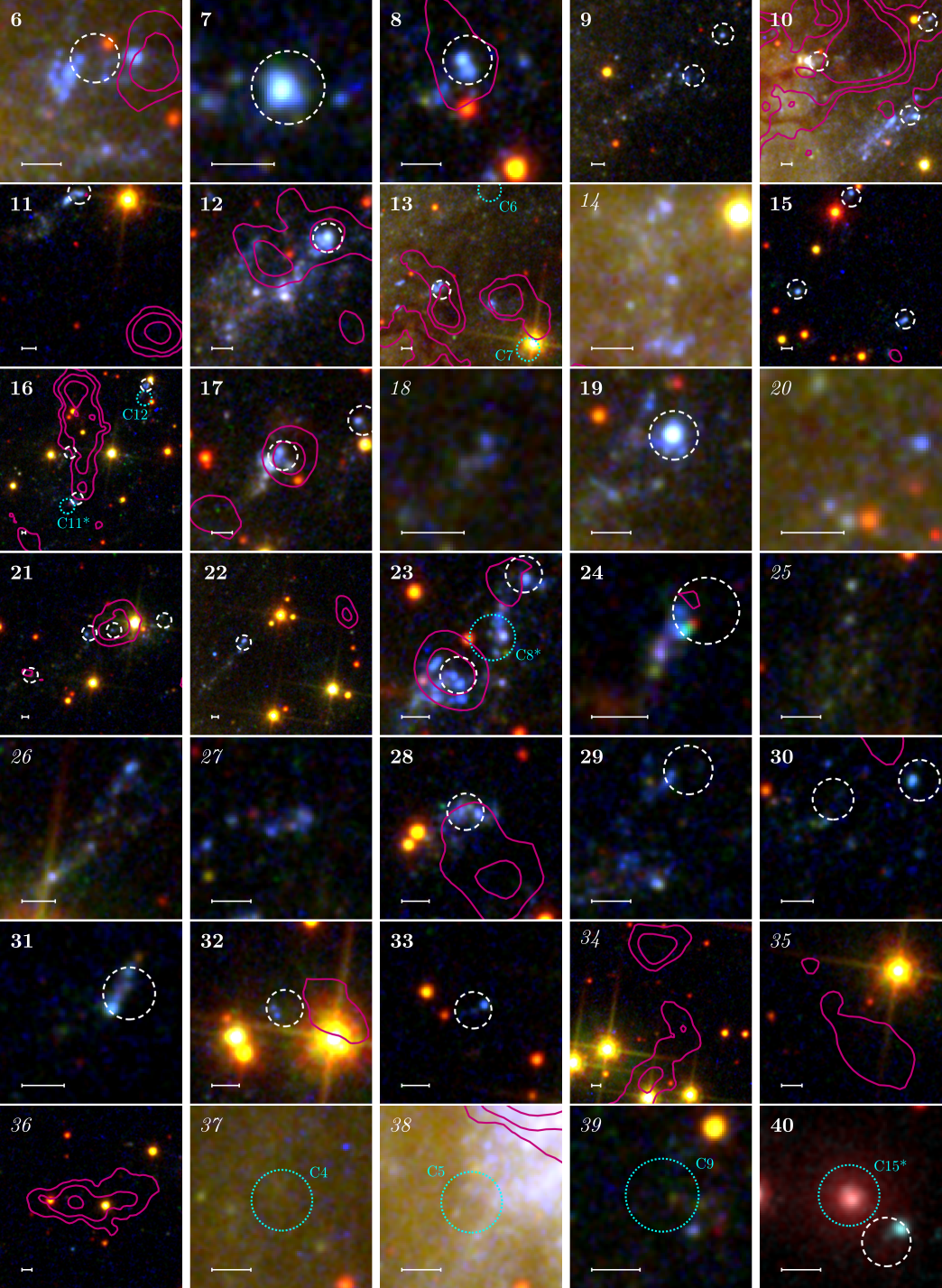}
    \vspace{-5pt}
    \contcaption{More zoom-ins.}
    \label{fig:wholeWithRegs2}
\end{figure*}

\eso{} (see Fig.~\ref{fig:wholeWithRegs}), a spiral galaxy near the center of the closest rich cluster,
the Norma cluster (Abell~3627), was first discovered as a galaxy undergoing \gls{rps} by \citet[hereafter \citetalias{sun2006}]{sun2006} with the \chandra{} and \emph{XMM} X-ray data that reveal a~70\,kpc long, narrow X-ray tail.
\citet[hereafter \citetalias{sun2007HA}]{sun2007HA} took the narrow-band imaging data with the \gls{soar} telescope to discover a~40\,kpc H$\alpha$ tail that aligned with the X-ray tail. More than~30 \ion{H}{ii} regions were also revealed, unambiguously confirming \gls{sf} in the stripped \gls{ism}. 
\citet[hereafter \citetalias{sun2010}]{sun2010} further studied \eso{} with deeper \chandra{} observations and \emph{Gemini} spectroscopic observations. The deep \chandra{} data surprisingly revealed that the X-ray tail, now detected to at least~80\,kpc from the galaxy, is bifurcated with a secondary branch on the south of the primary X-ray tail. Re-examination of the \citetalias{sun2007HA} H$\alpha$ data also shows H$\alpha$ enhancement at several positions of the secondary tail. \citetalias{sun2010} also presented the \emph{Gemini} spectra of over 33 \ion{H}{ii} regions and revealed the imprint of the galactic rotation pattern in the tail.

\citet{siv2010} presented \emph{Spitzer} data on \eso{} that reveals a 20\,kpc warm (130-160\,K) H$_{2}$ tail which is co-aligned with the X-ray and the H$\alpha$ tail (note that the extent of the H$_{2}$ tail is limited by the field of view of the \emph{IRS} instrument). The large H$_{2}$ line to IR continuum luminosity ratio suggests that \gls{sf} is not the main excitation source in the tail.
\citet{jachym2014} used \emph{APEX} telescope to observe \eso{} and its tail at four positions. Strong CO(2-1) emission is detected at each position, including the middle of the primary tail that is~40\,kpc from the galaxy, indicating abundant cold molecular gas in the tail. On the other hand, the \gls{sf} efficiency in the tail appears to be low.
\citet{jachym2019} further observed \eso{} and its tail with a mosiac from \gls{alma}
to obtain a high-resolution view of the cold molecular gas. The resulting map reveals a rich amount of molecular cloud structure in the tail, ranging from compact clumps associated with \ion{H}{ii} regions, large clumps not closely associated with any \ion{H}{ii} regions, to long filaments away from any \gls{sf} regions. 

Spatially resolved studies of warm, ionized gas have been revolutionized with the \gls{muse} on \emph{VLT}. As the first science paper with \gls{muse}, \citet{fumagalli2014} presented results from the \gls{muse} observations on \eso{} and the front half of the primary tail. The work demonstrates the great potential of \gls{muse} on studies of diffuse warm, ionized gas as often seen in stripped tails. The early \gls{muse} velocity and velocity dispersion maps suggest that turbulence begins to be dominant at $>$ 6.5\,Myr after stripping.
\citet[hereafter \citetalias{fossati2016}]{fossati2016}, with the same \gls{muse} data that \cite{fumagalli2014} used, presented a detailed study of line diagnostics in \eso{}'s tail. Their results also called for better modeling of ionization mechanisms in stripped tails. \citet{sun2021} presented a new \gls{muse} mosaic to cover the full extent of \eso{}'s tail, revealing very good correlation with the X-ray emission. This new set of \gls{muse} data provides a great amount of data for detailed studies of kinematics and line diagnostics \citep{Luo22}.

As discussed above, \eso{} has become the \gls{rps} galaxy with the richest amount of supporting data and multi-wavelength analysis.
In this context, a detailed study with the \hst{} data adds important information on this multi-wavelength campaign, which is the focus of this paper. 
Table~\ref{tab:esoProps} summarizes the properties of \eso{}.
As in \citetalias{sun2010}, we adopt a luminosity
distance of~69.6\,Mpc for Abell~3627 and $1\arcsec{} = 0.327$~kpc.
The rest of the paper proceeds as follows:
Section~\ref{sec:hstObs} details the \gls{hst} observations and the data reduction.
Section~\ref{sec:eso} focuses on the properties of \eso{}.
Section~\ref{sec:youngClusters} presents the studies of the young star complexes discovered in the tail of \eso{}.
Section~\ref{sec:discuss} is the discussion and
we present our conclusions in Section~\ref{sec:conclude}.
\input{sections/s1-intro/subs/eso-PropertiesTable}

%% file: sections/s1-intro/subs/eso-PropertiesTable.tex
\begin{table}
    \begin{center}
        \caption{Properties of ESO 137-001}
        \label{tab:esoProps}
        \begin{tabular}{c|c}
            \hline
            \hline
            Heliocentric velocity (\text{km/s})$^\mathrm{a}$    & 4647 (-224)  \\
            Offset (kpc)$^\mathrm{b}$     & 180     \\
            Position Angle            & $\sim$ 9$^\circ{}$ \\
            Inclination & $\sim$ 66$^\circ{}$ \\
            $W1$ (Vega mag)$^\mathrm{c}$  &  12.31 \\
            $L_{\rm W1}$ ($10^{9} L_{\odot}$)$^\mathrm{c}$ & 0.489 \\
            $W1 - W4$ (Vega mag) $^\mathrm{c}$      &  6.99  \\
            $m_{\rm F160W}$ (AB mag)$^\mathrm{d}$ & 13.16 \\
            Half light semi-major axis (kpc)$^\mathrm{d}$ & 4.91 \\
            M$_{\star}$ ($10^{9} M_{\odot})^\mathrm{e}$         & 5-8        \\
            $M_{\rm mol}$ ($10^9 M_\odot$)$^\mathrm{f}$ & $\sim$ 1.1 \\
            L$_{\rm FIR}$ ($10^{9} L_{\odot})$$^\mathrm{g}$   &  5.2 \\
            SFR (M$_{\odot}$/yr) (Galaxy)$^\mathrm{h}$ & 1.2 \\
            M$_{\star}$ ($10^{6} M_{\odot})$ (Tail)     & 2.5 - 3.0       \\
            SFR (M$_{\odot}$/yr) (Tail) & 0.4 - 0.65 \\
            Tail length (kpc)$^\mathrm{i}$ & 80 - 87 (X-ray/H$\alpha$) \\
            \hline
        \end{tabular}
    \end{center}

    Note: \\
    $^\mathrm{(a)}$ The heliocentric velocity of the galaxy from our study on the stellar spectrum around the nucleus (Luo et al. 2022). The velocity value in parentheses is the radial velocity relative to that of Abell~3627 \citep{woudt2004}.\\
    $^\mathrm{(b)}$ The projected offset of the galaxy from the X-ray center of A3627 \\
    $^\mathrm{(c)}$ The \emph{WISE} 3.4 $\mu$m magnitude, luminosity and the \emph{WISE} 3.4 $\mu$m - 22 $\mu$m color. The Galactic extinction was corrected with the relation from \citet{indeb2005}.\\
    $^\mathrm{(d)}$ The total magnitude and the half light semi-major axis at the F160W band.
    The axis ratio is 2.27 and the positional angle is $9.0^\circ$ east of the North.\\
    $^\mathrm{(e)}$ The total stellar mass estimated from \citetalias{sun2010}.\\
    $^\mathrm{(f)}$ The total amount of the molecular gas detected in the galaxy from \cite{jachym2014}.\\
    $^\mathrm{(g)}$ The total FIR luminosity from the \emph{Herschel} data (see Section~\ref{sub:galSFH})\\
    $^\mathrm{(h)}$ The average value from the first estimate (0.97) based on the \emph{Galex} NUV flux density and the total FIR luminosity from \emph{Herschel} with the relation from \citet{hao2011}, and the second estimate (1.4) based on the \emph{WISE} 22 $\mu$m flux density with the relation from \citet{lee2013}. The Kroupa \acrfull{imf} is assumed.\\
    $^\mathrm{(i)}$ The tail length for ESO 137-001 from~\citetalias{sun2010}.
    \label{tab:parameter}
\end{table}

%% file: sections/s2-hst/hst.tex
\section{\hst{} Observations and Data Analysis}\label{sec:hstObs}

The data for this work were collected using the \gls{wfc3} and \gls{acs}
on \gls{hst}, from proposals 11683 and 12372
(PI: Ming Sun). Details of the observations are summarized in Table~\ref{tab:hstObs}.
All observations focus on the tail of \eso{} described in \citet{sun2006,sun2007HA}.
The F275W data are especially sensitive to recent star formation in the last several~$10^7$ years and the F275W - F475W color measures the strength of the~4000\,\AA{} break.
The data from F475W and F814W add a broad spectrum of light to identify dim, diffuse
features, while the F160W data present the view of the galaxy least affected by the dust extinction.
Together, these four filters can be used to constrain the
stellar age of star clusters found in the galaxy and the tail.
The \gls{hst} \gls{wfc3} UVIS Channel has a pixel scale
of~0\farcs{}04 per pixel~(13\,pc/pix) with a \gls{psf} \gls{fwhm} of~0\farcs{}067
to~0\farcs089 \citep[22\,pc to~29\,pc;][]{dressel2022}.
The \gls{hst} \gls{wfc3} IR Channel has a pixel scale
of~0\farcs{}13 per pixel~(43\,pc/pix) with a \gls{psf} \gls{fwhm} of~0\farcs{}124
to~0\farcs156 \citep[41\,pc to~51\,pc;][]{dressel2022}.
The \gls{hst} \gls{acs} Wide Field Channel has a pixel scale
of~0\farcs{}05 per pixel~(16\,pc/pix) with a recoverable reconstructed \gls{fwhm} of~0\farcs{}100
to~0\farcs140~(33\,pc to~46\,pc) after dithering \citep{ryon2022}.

\input{sections/s2-hst/hst-DataTable}

Abell~3627 lies near the galactic plane which means the Milky Way extinction and foreground
clutter is high:~1.158 mag for F275W,~0.689 mag for F475W,~0.332 mag for F814W, and~0.108 mag for 
F160W~\citep{schlafly2011}.
We estimate the point source detection limits (as 3-$\sigma$ of the background rms) as~29.2\,mag
for F275W,~30.6\,mag for F475W,~29.9\,mag for F814W, and~31.3\,mag for F160W
(all corrected for the Galactic extinction).

The first step toward photometric measurements of \eso{} involved aligning the
images to each other.
We choose to perform this task using the STScI software tweakreg~%
\citep{gonzaga2012,avila2015} which is a member of the DrizzlePac software that
replaces the IRAF tweakshifts software.
We chose to align each image to
the distortion corrected image
created with the F814W \gls{hst} filter.
This band had a good signal-to-noise ratio as well as strong sources that
appeared in the other three bands.
Since there was no common misalignment across the four bands, the images in
each band had to be aligned to F814W on a per-band basis.
Absolute astrometry was also performed on the images by aligning to the
Guide Star Catalog~II~\citep[GSC2;][]{lasker2008}.
Overall, using this method, we were able to align the images to within~0\farcs{}01.

The next step in preparing the images for photometry was to drizzle the files
in each band together.
As with the alignment process above, we used a program in the DrizzlePac~%
\citep{gonzaga2012} called astrodrizzle~\citep{koek2003}.
This software assisted in lowering the noise floor as well as removal of \glspl{cr}
where there was overlap between images.
Each of the four filters has a different pixel scale.
Therefore, using astrodrizzle we were able to combine the images with a new,
similar pixel scale of~0\farcs{}03~(9.81\,pc).
We left most of the settings in their default state with the exception of the
parameters that set the image size and orientation.
We also changed the output weight image to an \gls{ivm} to provide correct noise estimates in
SExtractor~\citep{bertin1996}.
The uncertainties measured by SExtractor were then corrected for correlated noise per the
method detailed in \citet{hoffmann2021}.

Since there are only 2 - 3 frames for each filter, there are many residual \glspl{cr} in the chip
gaps and edges.
For sources detected in the chip gaps and edges, we ran a correlation between three bands (F275W,
F475W and F814W). Sources that are detected in only one band are considered \glspl{cr}. Such
sources were also visually inspected to ensure correlation accuracy. Identified \glspl{cr} were
then removed from the drizzled images.

We also created \gls{rms} images from the \gls{ivm} images~\citep{hammer2010} mentioned above by
taking the square root of the inverse of the \gls{ivm} images for use as the
SExtractor \citep{bertin1996} weights.

%% file: sections/s2-hst/hst-DataTable.tex
\begin{table*}
    \centering
    \caption{\hst{} Observations (PI: Sun)}
    \label{tab:hstObs}
    \begin{tabular}{ccccccc}
        \hline
        Filter & Instrument & Mode & Dither$^\mathrm{a}$ & Date & Exp (sec) & Mean $\lambda$ / FWHM ($\text{\AA}$) \\
        \hline
        F275W & WFC3/UVIS & ACCUM & 3 (2.4$''$) & 02/08/2009 & $3\times978.0$ &  2719/418 \\
        F475W & ACS/WFC & ACCUM & 2 (3.01$''$) & 02/08/2009 & $2\times871.0$ & 4802/1437 \\
        F814W & ACS/WFC & ACCUM & 2 (3.01$''$) & 02/08/2009 & $2\times339.0$ & 8129/1856 \\
        F160W & WFC3/IR & MULTIACCUM & 3 (0.605$''$) & 17/07/2011 & $3\times499.2$ & 15436/2874 \\
        \hline
    \end{tabular}
    \begin{tablenotes}
      \item {\sl Note:} $^\mathrm{(a)}$ number of dither positions (and offset between each dither).
    \end{tablenotes}
\end{table*}

%% file: sections/s3-eso/eso.tex
\section{ESO 137-001}\label{sec:eso}

\input{sections/s3-eso/subs/morph}

\input{sections/s3-eso/subs/dust}

\input{sections/s3-eso/subs/nuc}

%% file: sections/s3-eso/subs/morph.tex
\subsection{Morphology \& Light Profiles}
\label{sub:morph}

\begin{figure*}
    \centering
    \includegraphics[width=0.9\textwidth]{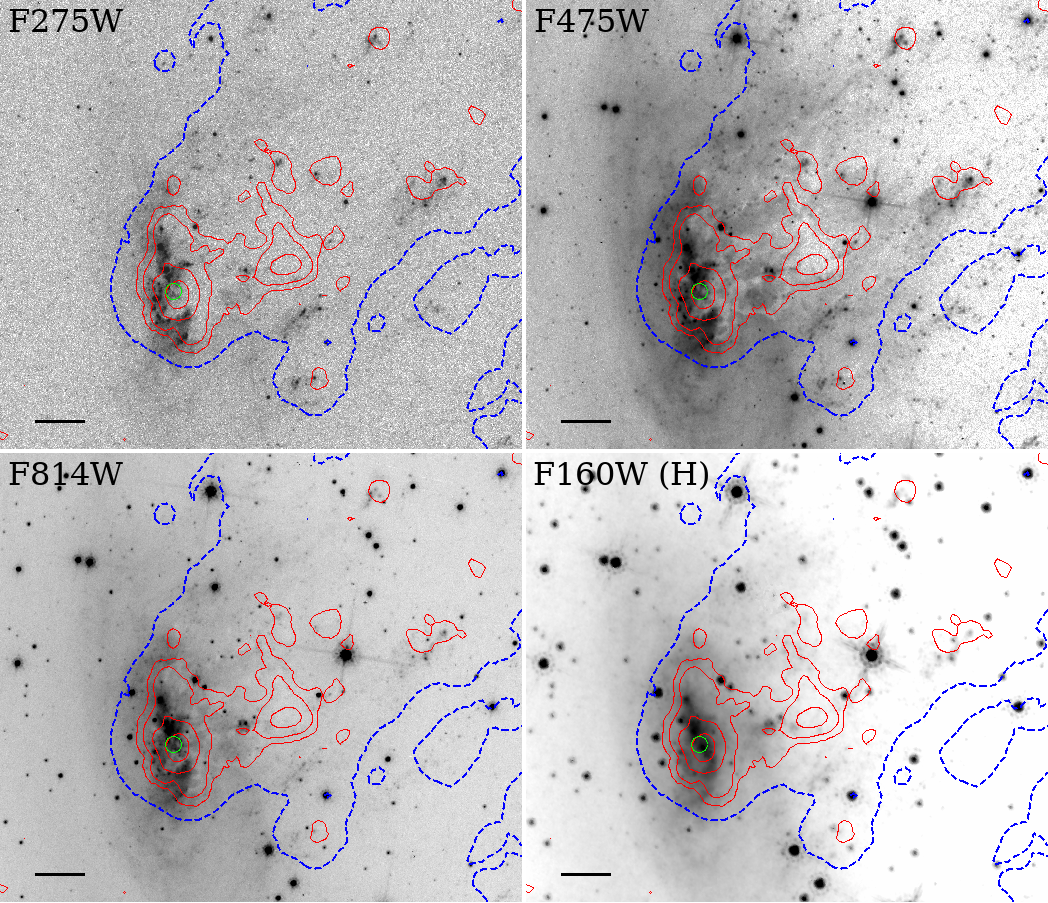}
    \vspace{-0.2cm}
    \caption{
    The central regions of the galaxy in the four \gls{hst} bands.
    One can see strong dust features downstream of the galaxy
    (vs smooth light distribution upstream) and enhanced \gls{sf} in the central part of the galaxy.
    The dashed blue contour is the outer H$\alpha$ edge from the full \gls{muse} mosaic image
    \citep{sun2021}, which shows the stripping front in the galaxy.
    The red contours show the CO emission from \citet{jachym2019}.
    There is a large CO clump downstream of the galaxy, at $\sim$ 2.3 kpc from
    the nucleus where dust attenuation is significant (especially apparent in the
    F475W image) and lacks bright star clusters and strong H$\alpha$ emission.
    This downstream region may be the ``deadwater'' region, or the stagnant part of the wake that is close to the moving body.
    Other smaller CO clumps are typically around young star complexes.
    The green circle (with a
    radius of 0.5$''$) shows the nucleus. The scale bar is~1\,kpc.
    }
    \label{fig:esoImageAllFilt}
\end{figure*}

A composite RGB image of \eso{} is shown in Fig.~\ref{fig:wholeWithRegs} which
includes zoomed in areas of interesting regions that will be discussed later this paper. Fig.~\ref{fig:esoImageAllFilt}
presents the central galaxy region in each of the four \gls{hst} filters.
\eso{}'s upstream region is nearly devoid of dust from
\gls{rps} while dust trails extend from the nucleus into the downstream regions. This
suggests that \gls{rps} has nearly cleared out the eastern half of the galaxy (also the near side) but \gls{rps} is still occurring around the nucleus and western half.
Fig.~\ref{fig:esoImageAllFilt} also shows the outer H$\alpha$ contour from the
\gls{muse} data that reveal the current stripping front that is only $\sim$ 1.2 kpc from the nucleus.
One can also see a large dust feature downstream has an associated large molecular
cloud detected by \gls{alma}.

\begin{figure*}
    \centering
    \includegraphics[width=\textwidth]{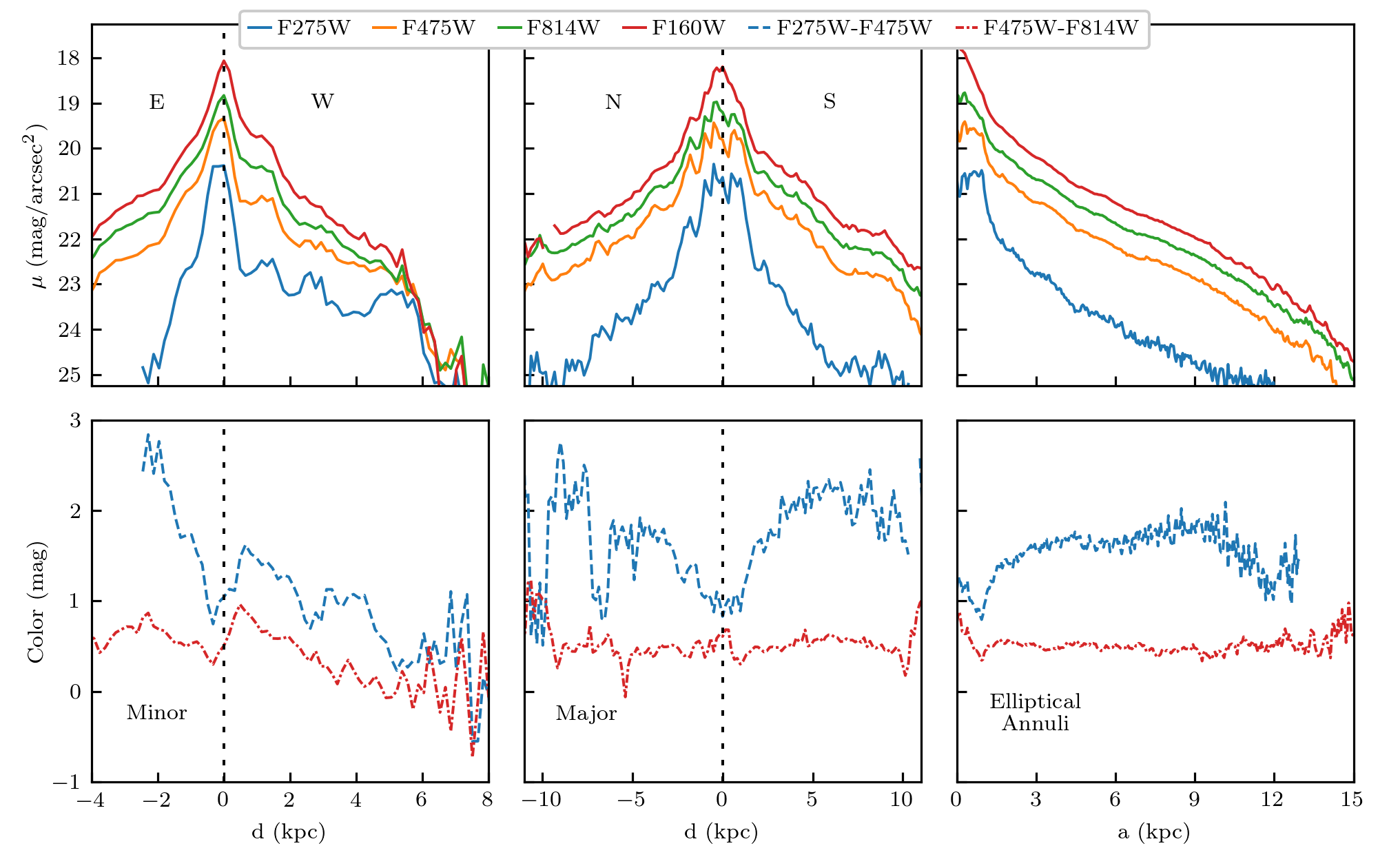}
    \vspace{-0.8cm}
    \caption{\emph{Top-Left:} The surface brightness profiles of \eso{} along its minor axis
    (a 6$''$ width and a 0.5$''$ step size). The vertical dashed line shows the position of the nucleus (also the same for other panels in this figure). The east (E) region
    is upstream because the major axis of the galaxy is almost NS.
    One can clearly see the upstream-downstream asymmetry especially in UV and blue bands.
    Note that the bin width is doubled for distances greater than~5\,kpc from the nucleus.
    \emph{Top-Middle:} The surface brightness profiles of \eso{} along its major axis (a 3$''$ width
    and a 0.5$''$ step size). One can see the light profiles along the major axis are much more
    symmetric than those along the minor axis. Note that the bin width is doubled for distances
    greater than~5\,kpc from the nucleus.
    \emph{Top-Right:} The radial surface brightness profiles of \eso{} in elliptical annuli (with
    the same proportions as the elliptical galaxy region in Fig.~\ref{fig:compRegions}).
    The $x$-axis represents the semi-major axis.
    \emph{Bottom Row:} The F275W - F475W and F475W - F814W profiles in the relevant regions from the
    upper panels.
    }
    \label{fig:esoProfile}
\end{figure*}

We quantitatively examine the galaxy structure by deriving the surface brightness profiles
in all four bands along the major axis, minor axis and in elliptical annuli centered at the
nucleus (Fig.~\ref{fig:esoProfile}). The galaxy center is set at the nuclear position defined
in Section~\ref{sub:nucleus}. The major axis has a position angle of~$9.0^\circ$ (measured
counterclockwise from North). The total F160W light measured from the galaxy is~13.16\,mag
(without correction from intrinsic extinction) and the half-light radius
is~$4.91 \pm 0.05$\,kpc in F160W.
This half-light radius in F160W can also be compared with the half-light radius of~4.74\,kpc
and~4.58\,kpc at $B$ band and $I$ band respectively from the \gls{soar} data \citepalias{sun2007HA}.
It is noteworthy that ESO~137\=/002 \citep{laudari2022} is~5.2 times brighter than \eso{}.
However, due to ESO~137\=/002's bright, central bulge, the half light size of ESO~137\=/002 is
approximately~2/3 the size of \eso{}.

The light profiles along the minor axis are shown in Fig.~\ref{fig:esoProfile}.
The profiles are measured in a series of rectangle regions each with a width of 40$''$ and a step size of
0.5$''$. Likewise, the profiles along the major axis are also shown. The profiles are measured in a
series of
rectangular regions each with a width of 20$''$ and a step size of 0.5$''$.
The elliptical profiles are measured in annuli with a width of 0.15$''$ along the major axis and have
an axis ratio of 0.44.
It is worth noting the asymmetry of the F275W profile along the minor axis
(Fig.~\ref{fig:esoProfile}, \emph{top-left}). The profile
shows the faintness of the ultraviolet light in the galaxy upstream regions and an excess of ultraviolet
light in the downstream. Since dust in the galaxy is mainly in the downstream region, the intrinsic
E-W contrast on the F275W - F475W color is in fact larger than shown in Fig.~\ref{fig:esoProfile}.
This demonstrates the quenched \gls{sf} upstream and enhanced \gls{sf} downstream in the near tail.
The radial profile also shows an F275W-F475W color gradient (Fig.~\ref{fig:esoProfile},
\emph{bottom-right}) where the galaxy is bluer in the central regions than it is in the
outer regions.
The F475W-F814W color profiles (Fig.~\ref{fig:esoProfile}, \emph{bottom row}) shows little change
along each axis in these wide bins.

We also derive the structural parameters of \eso{} in the F160W band
(least affected by intrinsic extinction)
using the two dimensional fitting algorithm GALFIT \citep{peng2002}.
For this case, we used a single S\'{e}rsic model, as well as a double component
model (bulge fitted by a S\'{e}rsic model while disk fitted by an
exponential component) to fit the galaxy image. The fitted parameters are listed in Table~\ref{tab:galfit}.
In the case of a single S\'{e}rsic model, there is a degeneracy between the
S\'{e}rsic index and the effective radius.
A double component model also fits the F160W image reasonably well. However, as shown in Fig.~\ref{fig:wholeWithRegs} and Fig.~\ref{fig:esoImageAllFilt}, there is no clear evidence for the existence of a bulge. The fits also suggest that any bulge component, if it exists, must be small. 
We note that both the total F160W light and the half-light radius from profile fitting
(Table~\ref{tab:esoProps}) are similar
to the double S\'{e}rsic model results with GALFIT (Table~\ref{tab:galfit}).
Overall, the derived S\'{e}rsic indexes are in good agreement with results from large surveys like
GAMA~\citep[e.g.,][]{lange2015} for galaxies similar to \eso{}.
Based on the light profiles and the GALFIT results (Table~\ref{tab:galfit}), we
measure the inclination angle of \eso{} to be~66$^\circ$ with the classic Hubble formula (assuming a morphological
type of SBc), which is the same as the result from HyperLeda \citep{hyperleda2014}.
As the motion of \eso{} is towards the east and mostly on the plane of sky, we conclude that the near side of the galaxy is towards the east, as the stripped dust clouds need to be located between the disk and us the observer to make the downstream dust features significant.
Another way to conclude the east side as the near side is from the spiral arm winding (Fig.~\ref{fig:extinction}). As almost all spiral arms are trailing, \eso{}'s spiral arms are rotating counter-clockwise. As the south side of the galaxy is rotating away from us relative to the nucleus \citep{fumagalli2014}, the east side must be the near side.

\input{sections/s3-eso/subs/galfit-table}

%% file: sections/s3-eso/subs/galfit-table.tex
\begin{table}
    \begin{center}
    \caption{The GALFIT fits results for F160W image of \eso{}}
    \label{tab:galfit}
    \vspace{-0.1 cm}
    \begin{tabular}{cccc}
        \hline
        \hline
        Parameter & Single & Double  \\
                  & ($\chi_{\nu}^{2}$=5.323/6.293) & ($\chi_{\nu}^{2}$=5.206)\\
        \cline{3-4}
        &  & Bulge-S\'{e}rsic & Disk-Exp \\
        \hline
        Total mag                     & 12.68/13.78 & 15.77  & 13.34  \\
        $r_e$ (kpc)                   & 9.78/2.74 & 0.91  & 5.52\\
        S\'{e}rsic index              & 2.97/(1.0) & 1.02  & (1.0) \\
        Axis ratio                    & 0.436/0.469 & 0.351  & 0.468 \\
        PA ($\deg{}$)                 & 8.81/8.22 & 14.3 & 5.59 \\
        \hline
        \hline
    \end{tabular}
    \end{center}
    \vspace{-0.2cm}
    Note: The axis ratio is the ratio between the minor axis and the major axis. The position angle is measured relative to the north and counter-clockwise. For the single S\'{e}rsic model, the fit with the index fixed at 1.0 (for an exponential disk) is also shown. Parameters in parentheses are fixed.
\end{table}

%% file: sections/s3-eso/subs/dust.tex
\subsection{Dust Features}
\label{sub:dust}

\begin{figure*}
    \includegraphics[width=\textwidth]{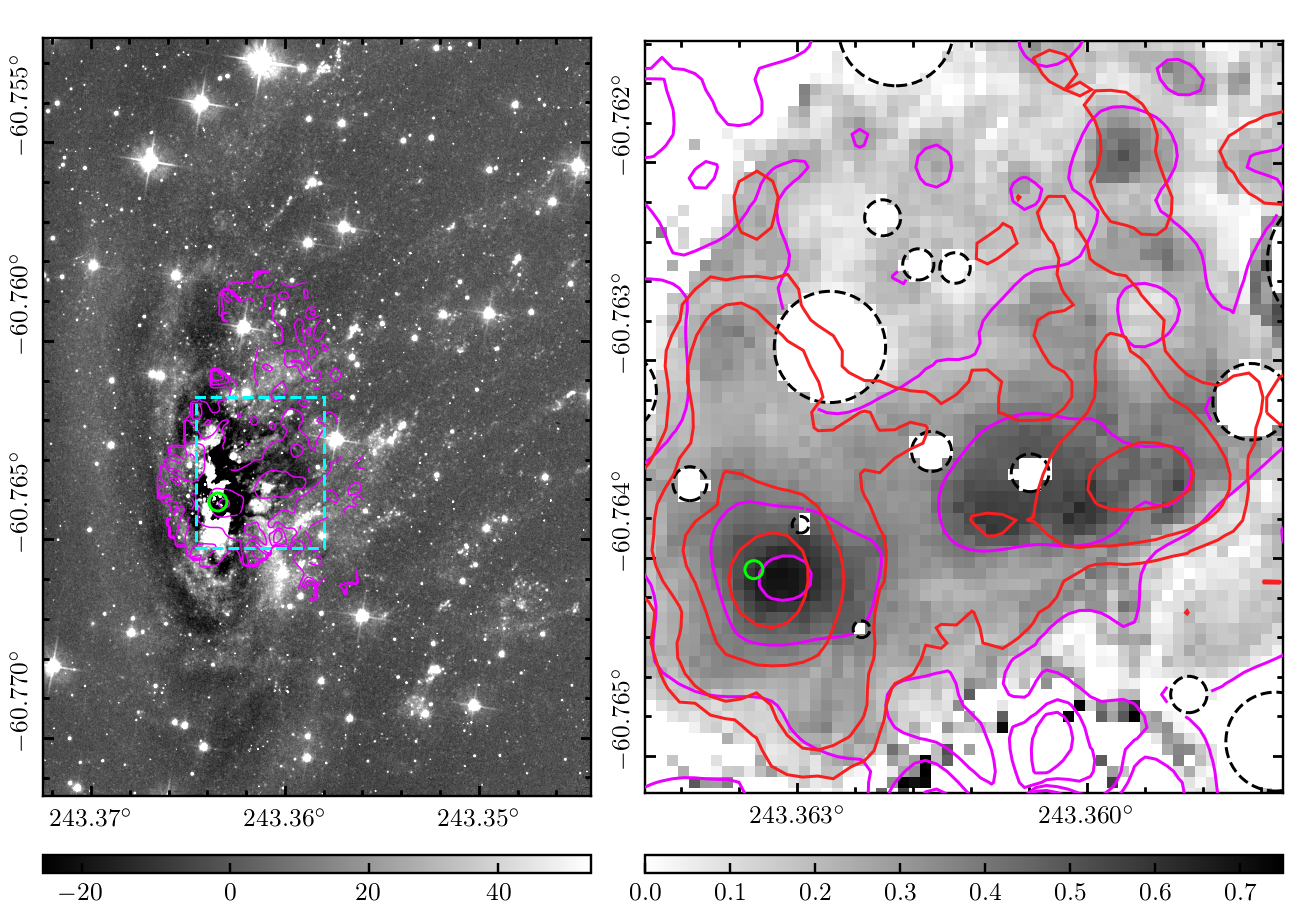}
    \vspace{-0.6cm}
    \caption{\emph{Left:} The residual F475W image of \eso{}, after subtracting the one-S\'{e}rsic model fit obtained with GALFIT (Table~\ref{tab:galfit}). Spiral arms, dust features downstream of the galaxy and some blue streams composed of young stars can be seen. The galactic nucleus (see Section~\ref{sub:nucleus}) is marked with the small green circle in both panels.
    The footprint of the E(B-V) image on the right is indicated by the cyan rectangle.
    \emph{Right:} The E(B-V) map around \eso{}, derived from the Balmer decrement with the \gls{muse} data (see Section~\ref{sub:dust}), assuming the \citet{calzetti2000} extinction law.
    The magenta contours (present in both panels) represent the derived E(B-V) values of~0.0,~0.2,~0.4, and~0.6.
    The red contours show the CO emission from \citet{jachym2019} as shown in
    Fig.~\ref{fig:esoImageAllFilt}.
    Bright foreground stars were masked in dashed-line circles. One can see a general good correlation between dust features, high extinction regions and CO clouds.
    }
    \label{fig:extinction}
\end{figure*}

To better show the dust features in the galaxy, we also used GALFIT to produce a residual
image with a single S\'{e}rsic model in the F475W image. Prior to the analysis, foreground stars
are masked. The residual image is shown in Fig.~\ref{fig:extinction}. It shows some prominent
dust features downstream of the galaxy (also see Fig.~\ref{fig:esoImageAllFilt}). The spiral
pattern of the galaxy is also better shown after a smoothed component removed. Some dust features downstream of the galaxy can also be seen clearly.

As \gls{muse} has fully covered the galaxy and the tail \citep{Luo22}, we can use the Balmer decrment to constrain the intrinsic extinction. Our analysis is similar to the work done in \cite{fossati2016} but with the new data. The stellar spectrum has also been subtracted in the analysis (see \citealt{Luo22} for detail). The \citet{calzetti2000} extinction law is assumed.
The resulting E(B-V) map close to the galaxy is shown in Fig.~\ref{fig:extinction}-\emph{right}.
The map shows a region of high extinction a few arcseconds west of the galaxy nucleus with
an additional area of high extinction near the CO clump shown by the \gls{alma} data \citep{jachym2019}.
The \gls{muse} extinction map also shows reasonable correlation to the GALFIT residual image in Fig.~\ref{fig:extinction}-\emph{left}.

%% file: sections/s3-eso/subs/nuc.tex
\subsection{Nucleus}
\label{sub:nucleus}

There is no evidence for an \gls{agn} in ESO 137-001 from X-ray, optical, NIR/MIR and radio \citep{sun2007HA,sun2010,siv2010,fossati2016}, which makes the identification of its nuclear region not straightforward. We constrain the position of the nucleus from the peaks of the X-ray, H$\alpha$, CO emission in the galaxy \citep{sun2010,fossati2016,jachym2019}, and the GALFIT fits to the \gls{hst} images, as all these peak positions are consistent with each other within $\sim 1''$. The nucleus position is then determined by averaging these positions at (16:13:27.231, -60:45:50.60) with an uncertainty of~0\farcs{}5. There is strong \gls{sf} ongoing around the nucleus and the nuclear region still retains a lot of molecular gas.

%% file: sections/s4-youngclusters/h2_regs.tex
\section{Young Star Complexes in the Tail}\label{sec:youngClusters}

\input{sections/s4-youngclusters/subs/regions}

\input{sections/s4-youngclusters/subs/color}

\input{sections/s4-youngclusters/subs/ssp_comp}

\input{sections/s4-youngclusters/subs/props}

\input{sections/s4-youngclusters/subs/muse_alma}

\input{sections/s4-youngclusters/subs/chandra}

%% file: sections/s4-youngclusters/subs/regions.tex
\subsection{Regions of Interest and Source Sample}\label{sub:roi}

\begin{figure}
    \vspace{-0.3cm}
    \includegraphics[width=\columnwidth]{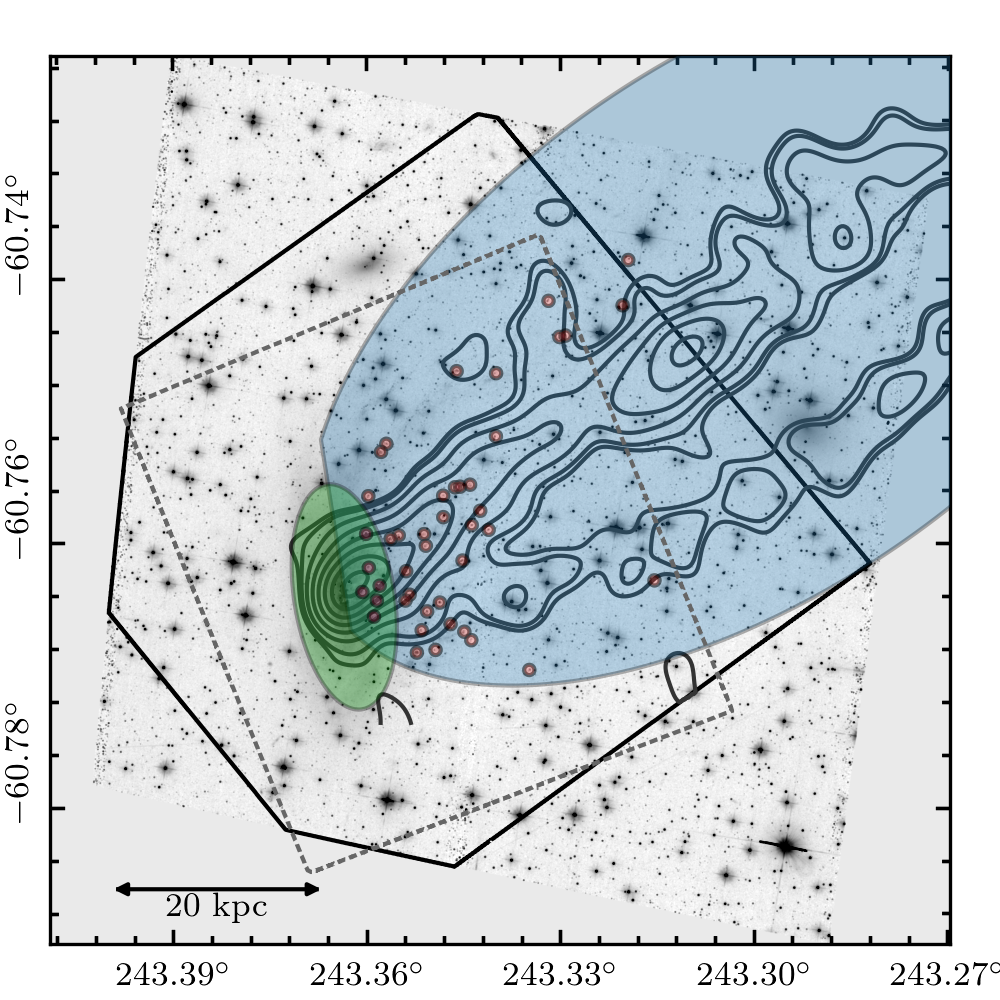}
    \vspace{-0.5cm}
    \caption{
    Regions of interest to be studied in this paper are shown here using the F475W
    image as a reference. The small, red circles denote the 43 \ion{H}{II} regions identified
    from the full \gls{muse} data (see Section~\ref{sub:roi}).
    The green ellipse centered at the nucleus defined in Section~\ref{sub:nucleus},
    angled at~$9.0^\circ$ from the North and with with semi-axes ($31\farcs{}00$, $13\farcs{}64$),
    defines the galaxy region. The galaxy region also excludes the six \ion{H}{ii} regions
    that fall in the green ellipse. We also measure the photometry of the galaxy
    region in~14 smaller regions in~7 radial bins, with each bin divided into the upstream
    and downstream portions.
    The largest, blue ellipse indicates the tail regions and is defined
    by an ellipse centered at (16:13:12.1968, -60:44:36.495) rotated
    to~$34.7^\circ$ above the RA axis with axes ($137\farcs{}137$, $72\farcs{}560$).
    The exception to this tail region is that we exclude the upstream sources
    which are defined as anything to the east of the major axis of the galaxy,
    as well as the galaxy region and \ion{H}{II} regions.
    The control region is therefore defined as the area outside the tail and galaxy
    regions. Note that our analysis is limited to the field covered by F275W, F475W
    and F814W observations together, which is marked by the solid, black line. The smaller
    dotted, gray line indicates the WFC3/F160W \gls{fov}. Finally, the gray contours show
    the \chandra{} X-ray emission reported in \citet{sun2010}. Only the near part of
    the~$\sim$80\,kpc X-ray tail is covered by \gls{hst}.
    }
    \label{fig:compRegions}
\end{figure}

While \citetalias{sun2007HA} defined a sample of \ion{H}{II} regions in the tail of
\eso{} from the \gls{soar} narrow-band imaging data, the recent full \gls{muse}
mosaic \citep[more detail in][]{sun2021,Luo22} provides much better
data to select \ion{H}{II} regions.
We selected \ion{H}{II} regions from the extinction-corrected \gls{muse} H$\alpha$
surface brightness map with SExtractor.
Sixty-four candidates are identified by selecting for CLASS\_STAR~$>$~0.8 (point-like sources)
and the ellipticity~($e$)~$<$~0.55.
We relaxed the criteria on $e$ (\citealt{fossati2016} used $e < 0.2$) as several \ion{H}{II} regions are mixed with the stripped H$\alpha$ filaments, which will enhance the ellipticity obtained by SExtractor. We further applied a limit for the integrated H$\alpha$ flux of the candidates
as~$2.1 \times 10^{-16}$ cm$^{-2}$ erg s$^{-1}$ to avoid the selection of faint H$\alpha$ clumps in the tail.
In addition, the {\mbox{[N\,\textsc{ii}]}}/H$\alpha$ emission-line ratio was also required to be
less than~0.4 to confirm the ionization characteristic of the \ion{H}{II} candidates.
We finally selected~43 \ion{H}{II} regions in the stripped tail of \eso{}. As shown in Fig.~\ref{fig:compRegions},~42 of them are covered by the F275W data, while the other one is just off the F275W \gls{fov} but covered by the F475W/F814W data.
\citetalias{sun2007HA} presented a sample of~29 \ion{H}{II} regions, plus~6 more candidates.~27 of
these~29 sources are also selected by \gls{muse}.
The other two are also shown as compact H$\alpha$ sources in the \gls{muse} data.
They would have been selected with a lower flux limit than what was adopted.
For the~6 candidates in \citetalias{sun2007HA},~3 are \gls{muse} \ion{H}{II} regions.
Two others are also shown as compact H$\alpha$ sources but fainter than the chosen threshold.
One is not confirmed with the \gls{muse} data.
This comparison shows the robustness of the \ion{H}{II} regions selected in \citetalias{sun2007HA}.
The new \gls{muse} \ion{H}{II} region sample also adds~13 new \ion{H}{II} regions compared
with \citetalias{sun2007HA}. These new ones are typically fainter than \ion{H}{II} regions selected by \citetalias{sun2007HA} and they are generally close to bright stars. The \citetalias{sun2007HA} selection is essentially based on H$\alpha$ \gls{ew} so these faint ones that are close to bright stars were not included in \citetalias{sun2007HA}.
\cite{fossati2016} selected a sample of 33 \ion{H}{II} regions with the 2014 \gls{muse} data that has less coverage of the tail and are also shallower around the galaxy. Thirty two of them are also in the \ion{H}{II} sample of this work. The only one missing has a CLASS\_STAR of 0.76 with the now deeper data than what \cite{fossati2016} used, just below the threshold we adopted.

\begin{figure}
    \vspace{-0.1cm}
    \includegraphics[width=\columnwidth]{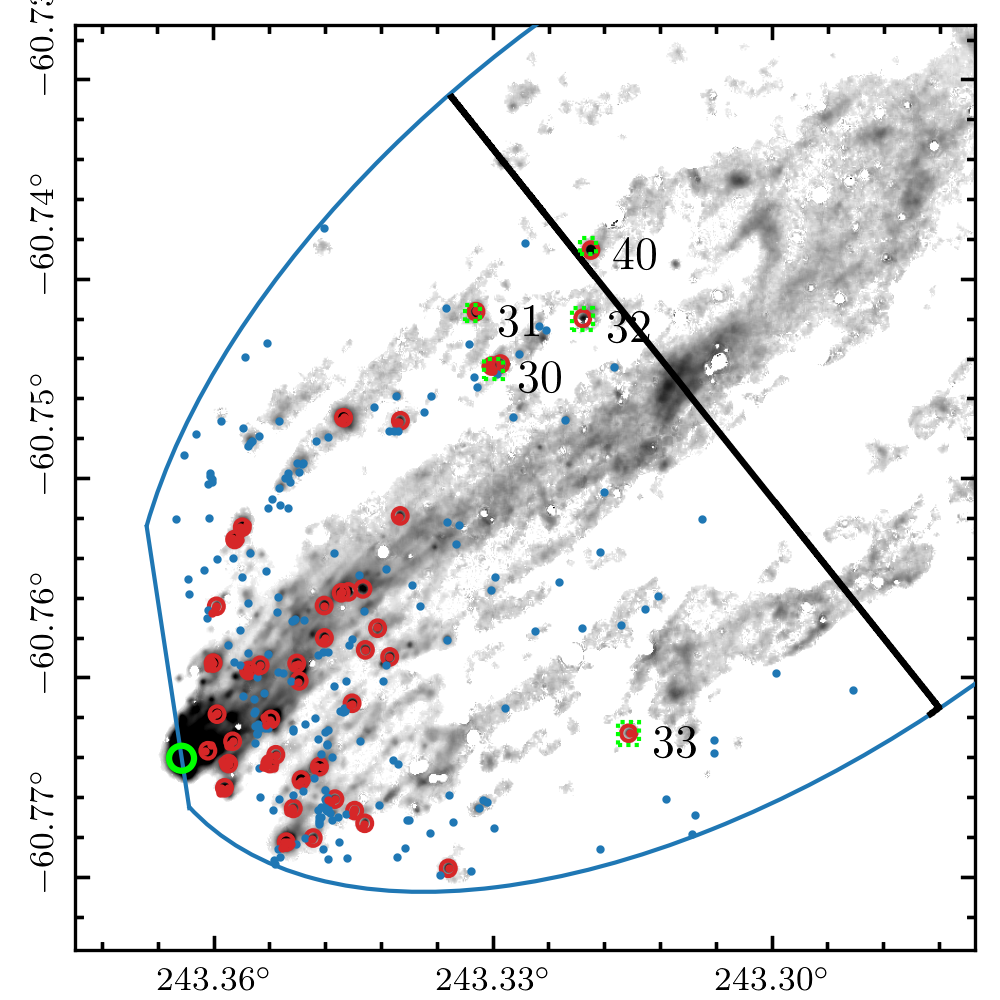}
    \vspace{-0.6cm}
    \caption{The \ion{H}{II} (red) and blue tail sources (blue) plotted in Fig.~\ref{fig:compRegions} are shown on the \gls{muse} H$\alpha$ image from \citet{sun2021}. The solid black line again indicates the field covered by F275W, F475W, and F814W, as shown in Fig.~\ref{fig:compRegions}. The galaxy nucleus is marked with a green circle on the left. The missing zoom-in regions from Fig.~\ref{fig:wholeWithRegs} are plotted using green dashed squares.
    }
    \label{fig:srcPositions}
\end{figure}

With the \ion{H}{ii} regions defined from the \gls{muse} data, we define four regions of
interest (Fig.~\ref{fig:compRegions}) for the subsequent analysis:
small red circles --- \gls{muse} \ion{H}{II} regions defined in this work
(each represented by a circle with a radius of 1\farcs{}4);
the green ellipse --- the galaxy region;
the large blue ellipse (but within the thick black line to show the common \gls{fov} of the
F275W, F475W and F814W data) --- the tail region;
the area outside of the green and blue ellipses but still within the thick black line --- the control region.
There is common area shared by different regions so it is defined that the galaxy region is the green ellipse excluding small red circles. The tail region is the blue ellipse (but within the
thick black line) excluding green ellipse and small red circles.
The sky areas for each of the four regions after removal of bright stars
are~0.065,~0.313,~3.123, and~2.001 arcmin$^2$ for the \ion{H}{II}, galaxy, tail, and
control regions, respectively.

For the \hst{} photometry studies of individual sources, we define the baseline
sample of sources as sources covered by the F275W, F475W, and F814W data.
The baseline source selection
is defined as follows.
First, spurious sources from wrong alignment, scattered light from bright stars and residual \glspl{cr}
close to the edges were removed. This leaves~3803,~12293 and~10783 sources in F275W, F475W and F814W
respectively with  SExtractor.
Second, only sources detected in at least two adjacent bands were kept. This includes sources
detected in all three bands, sources detected in F275W and F475W, and sources detected in F475W and
F814W.
From this selection,~713 were detected in all three bands,~144 were detected in F275W/F475W,
and~4882 were detected in F475W/F814W. After taking the union of these three sets,~5739 sources remain.
Third, stars in the the GSC2 catalog \citep{lasker2008} were removed. The number of sources was
dropped to~5422.
Fourth, sources that were brighter than~19.45\,mag in F475W,~20.40\,mag in F814W, and~21.1\,mag in F160W
were removed.
The F475W magnitude cuts are one magnitude brighter than the brightest star cluster in \eso{} and its tail
while the F814W and F160W cuts were chosen based on the color-magnitude diagram discussed in
section~\ref{sub:sfProps}.
Fifth, red sources with F275W-F475W~$>$~2.90 mag and F475W-F814W~$>$~2.00 mag are removed (see
Fig.~\ref{fig:colorComps} for the corresponding ages).
The above two steps decreased the source number to~908.
Sixth, sources with individual mag error greater than one mag were removed, which further decreased
the source number to~520.
This final sample includes~127 in the \ion{H}{II} regions,~201 in the tail,~140 in the galaxy and~36
in the control region.
The sources defined in the baseline sample are shown in
Fig.~\ref{fig:srcPositions} which presents the sources identified in the \ion{H}{II} regions in red and other sources identified within the tail.

%% file: sections/s4-youngclusters/subs/color.tex
\subsection{Color - Color diagram}
\label{sub:colorResults}

\begin{figure*}
    \includegraphics[width=0.75\textwidth]{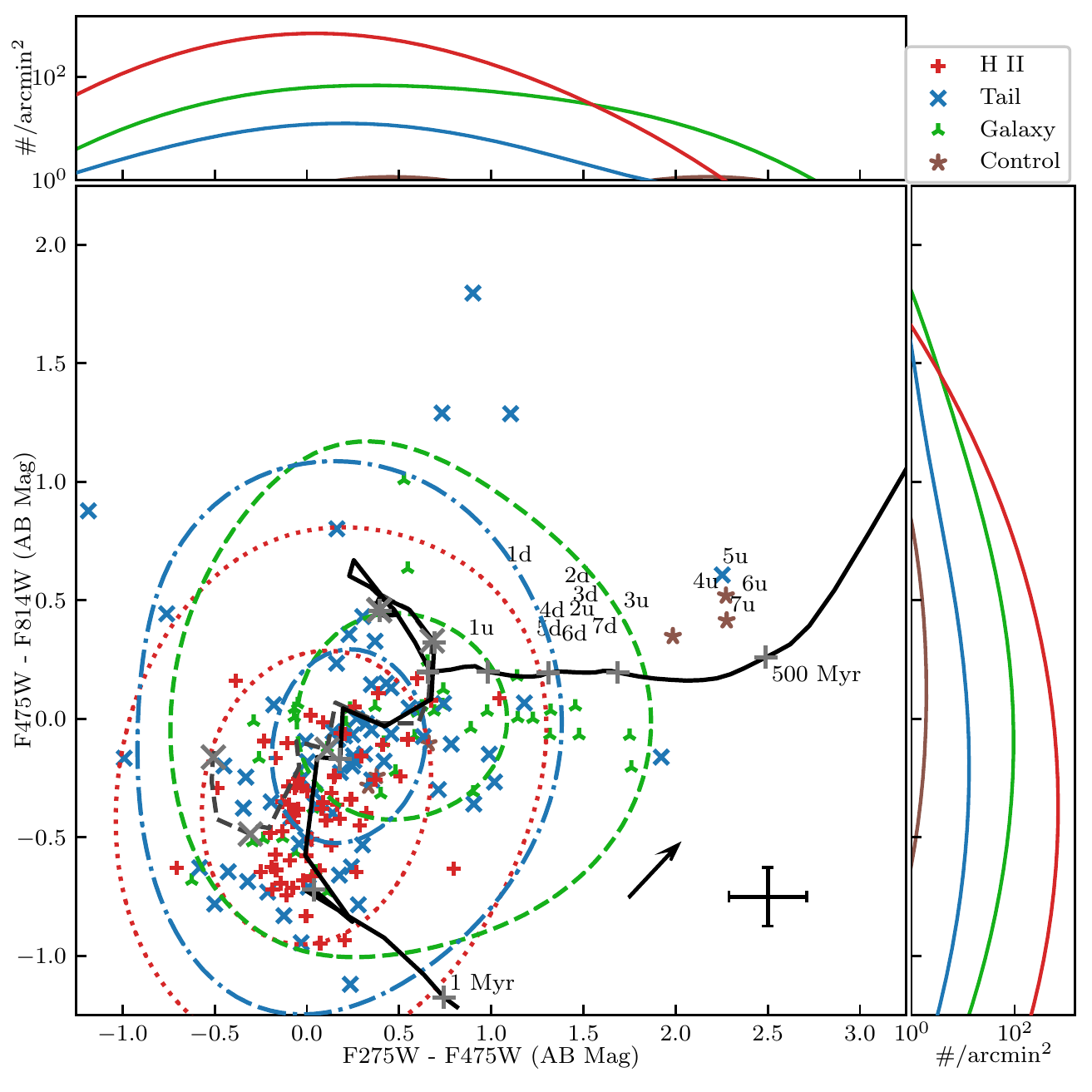}
    \vspace{-0.4cm}
    \caption{
    The colors of the \hst{} sources detected in the regions defined in Fig.~\ref{fig:compRegions}.
    Sources displayed here were detected in all three bands.
    Only sources with a photometry error of less than 1 AB
    magnitude in each filter were selected for studies here.
    The line plots on the top and right of the figure are the Gaussian \glspl{kde}
    ($\sigma=0.497$ AB mag which is median combined error of the data)
    of the three band detections
    that have been normalized by the search area of each region (see Section~\ref{sub:roi}).
    The displayed number of
    sources for the \ion{H}{II}, tail, galaxy, and control regions are~65,~67,~43
    and~6, respectively.
    The median error of all displayed sources is plotted in the bottom-right of the main figure.
    The text markers represent each of the
    seven radial upstream/downstream galaxy
    sub-regions defined
    in Fig.~\ref{fig:compRegions} where the number in each annotation
    represents the radial bin (``1'' at the nucleus) and the ``u'' and ``d''
    represent the upstream and downstream regions, respectively.
    The Starburst99 + Cloudy track
    (the solid, black line) is superposed.
    The Starburst99 track is the Genv00 model with $Z=0.014$.
    The ticks (which are more clearly labeled in Fig.~\ref{fig:colorCompsPlusLimits}) on each track represent
    ages of~1,~3,~5,~7,~10,~30,~50,~100,~200, and~500\,Myr~%
    (1\,Myr with the bluest F475W - F814W color as marked and~500\,Myr at the other end).
    The gray dashed line at ages of less
    than~10\,Myr represents the track with the Starburst99 model only.
    An intrinsic extinction value of $E(B-V) = 0.08$ (see Section~\ref{sub:sspComp} for detail)
    has been applied to the tracks according to the \citet{calzetti2000} extinction law. The arrow on the figure shows this $E(B-V) = 0.08$ extinction effect on the track.
    We also show the~7.5\% and~25\% \gls{kde} contours for all sources in these three regions
    (detections plus limits, see Fig.~\ref{fig:colorCompsPlusLimits} for more details).
    }
    \label{fig:colorComps}
\end{figure*}

We present the color-color (F475W-F814W versus F275W-F475W) results of this baseline sample in Fig.~\ref{fig:colorComps} to determine the characteristics of the young star complexes in and around \eso{}.\footnote{We do not include the F160W data in the analysis of tail sources for its poorer angular resolution than other \hst{} bands and the general faintness of young star complexes at the NIR.}
The figure presents a large number of sources that meet the above criteria within the \ion{H}{II} and tail regions and, a smaller number of sources within the galaxy and control regions. However, the \glspl{kde}%
\footnote{A \gls{kde} is performed by placing a chosen kernel (e.g., a Gaussian) at every
data point then performing a sum over the set of kernels over all space.}
suggest that the galaxy has a higher number of sources per square arcminute than the tail.

\begin{figure*}
    \includegraphics[width=0.75\textwidth]{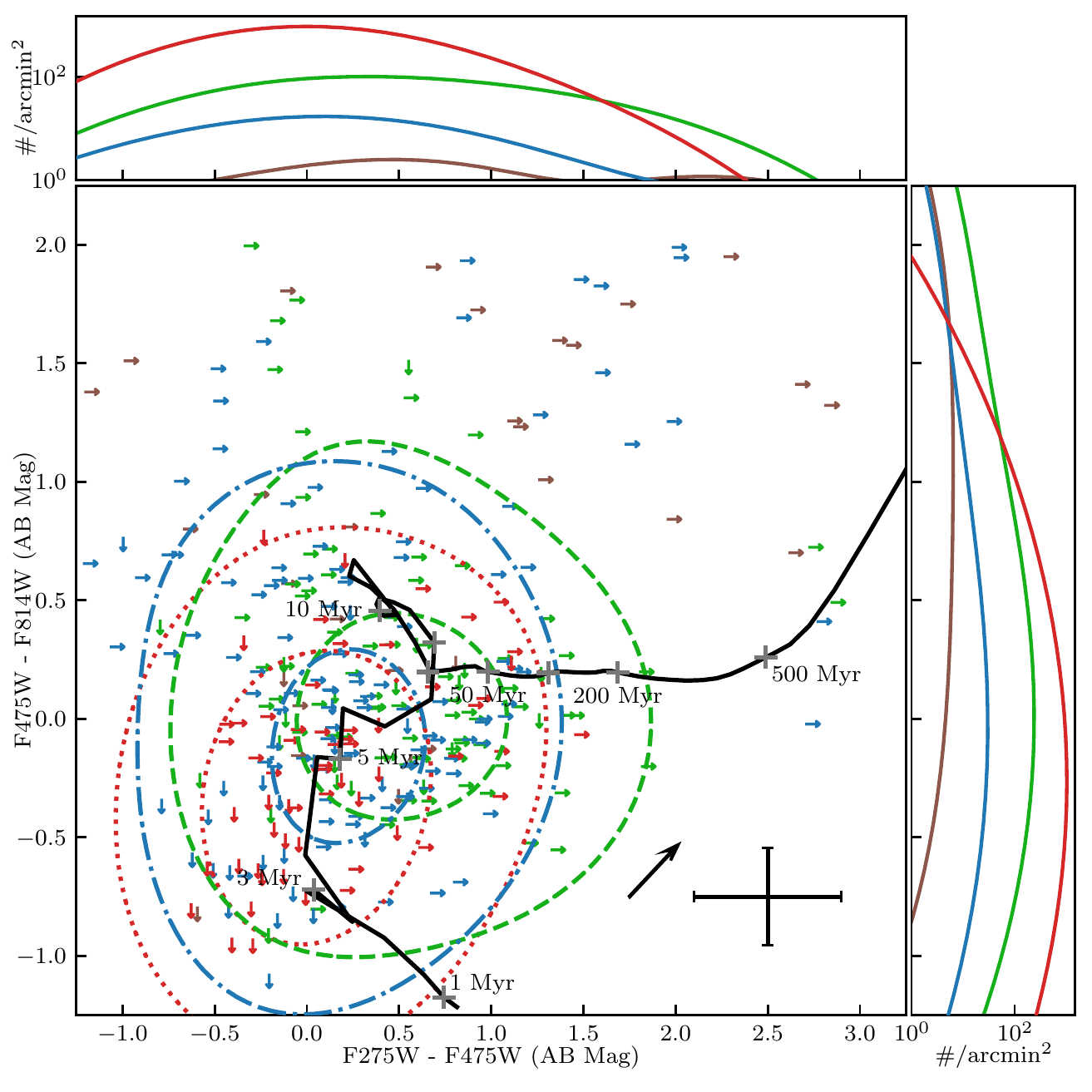}
    \vspace{-0.4cm}
    \caption{
    Similar to Fig.~\ref{fig:colorComps} but limits on colors are shown.
    Sources with a right arrow were detected in F475W and F814W but not F275W. Sources with a
    down arrow were detected in F275W and F475W but not in F814W. No genuine sources detected
    in F275W and F814W but not in F475W were identified.
    Otherwise the source selection is the same as Fig.~\ref{fig:colorComps}.
    We also show the same \gls{kde} contours for all sources (three band detections plus limits) as shown in Fig.~\ref{fig:colorComps}.
    The total number of
    sources for the \ion{H}{II}, tail, galaxy, and control regions are~127,~201,~140
    and~36, respectively.
    The F275W-F475W median errors for each region are~0.20,~0.47,~0.52, and~0.70 AB
    magnitudes, and the F475W-F814W median errors are~0.15,~0.26,~0.20, and~0.16 AB
    magnitudes, for \ion{H}{II}, tail, galaxy and control regions respectively.
    The median error of all sources is plotted in the bottom-right of the main figure.
    The track is the same Starburst99 + Cloudy track shown in Fig.~\ref{fig:colorComps}. The line plots on the top and right of the figure now show the Gaussian \glspl{kde} for detections plus limits.
    }
    \label{fig:colorCompsPlusLimits}
\end{figure*}

Fig.~\ref{fig:colorComps} also indicates that the majority of sources in the
\ion{H}{II} regions tend to be bluer than those in the galaxy.
While there is some overlap between the two sets, the two are distinct regions on the color-color diagram.
The sources in the \ion{H}{II} and tail regions are also bluer than the sources in the control region.
We also compared the \gls{kde} of the \ion{H}{II} and tail regions in Fig.~\ref{fig:colorComps}.
The high source number density in the \ion{H}{II} regions is mainly from its small area by definition.
After removing the background contribution estimated from the control region, there are only $\sim$ 18\%
more sources in the tail region than in the \ion{H}{II} regions and it is also found that sources in
the \ion{H}{II} and tail regions have very similar color distributions, which suggests them both as young star complexes.
We also show limits (or two-band detections) in Fig.~\ref{fig:colorCompsPlusLimits} as some of them can be low-mass young star clusters (lacking F814W detection) or older star clusters (lacking F275W detection) in the tail.

%% file: sections/s4-youngclusters/subs/ssp_comp.tex
\subsection{Comparison with SSP tracks}\label{sub:sspComp}

\begin{figure}
    \centering
    \vspace{-0.1cm}
    \includegraphics{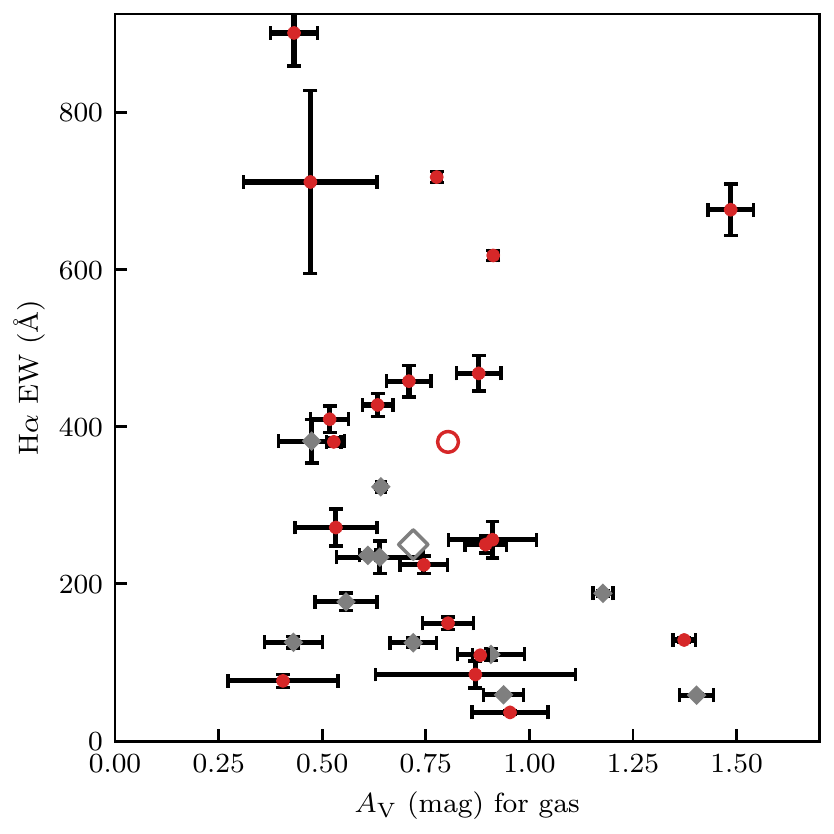}
    \vspace{-0.3cm}
    \caption{
    The H$\alpha$ \gls{ew} vs. the intrinsic extinction $A_\mathrm{V}$ for~33 \ion{H}{II} regions. To derive the H$\alpha$ \gls{ew}, we accounted for the different intrinsic extinction for gas and stars, E(B-V)$_{\mathrm{star}}$ / E(B-V)$_{\mathrm{gas}}$ = 0.44, as discussed in Section~\ref{sub:sspComp}.
    Sources with \gls{muse} stellar contamination or an unreliable H$\beta$ measurement were removed.
    The grey points are \ion{H}{ii} regions where bright stars are detected in \gls{hst} within 1.5$''$. The \gls{muse} spectra of these \ion{H}{ii} regions have elevated continuum from bright stars so the H$\alpha$ \gls{ew} is under-estimated, while the $A_V$ results are still valid. The~21 red points are \ion{H}{ii} regions without bright stars within 1.5$''$ in \gls{hst}.
    The large, open gray diamond shows the median values (EW=236\,\AA{} and $A_V$=0.78\,mag) for all~33 sources, while the large, open red circle shows the median values for red points (EW=381\,\AA{} and $A_V$=0.80\,mag).
    The median \gls{ew} values are~171\,\AA{} and~217\,\AA{}, respectively, when the~0.44 factor is not applied.
    }
    \label{fig:museEWvsAV}
\end{figure}

Fig.~\ref{fig:colorComps} and Fig.~\ref{fig:colorCompsPlusLimits} also compares the colors of the \gls{hst} sources to a Starburst99 \citep{leith1999} track. Particularly, the Genv00 model \citep{ekstrom2011} with an instantaneous burst, a \citet{kroupa2001} \gls{imf} and $Z = 0.014$ is assumed.
While in principle we can use the color-color relation in combination with the \gls{ssp}
track to constrain the age of young star complexes, intrinsic extinction in young star complexes
needs to be corrected first.
We can determine the intrinsic extinction in \ion{H}{ii} regions defined from the \gls{muse} data,
with the classic Balmer decrement method \citep[e.g.,][]{fossati2016}.
As shown in Fig.~\ref{fig:museEWvsAV}, the median $A_{\rm V}$ is 0.78 mag for~33 \ion{H}{ii} regions.
If the six regions in the galaxy region are excluded, the median $A_{\rm V}$ is 0.72 mag.
If the same extinction law as used in \cite{poggianti2019-g13} is instead used,
the $A_{\rm V}$ decreases by $\sim$ 9\%.
Thus, the median $A_{\rm V}$ of \ion{H}{ii} regions in \eso{}'s tail is comparable to the
median value of~0.5 mag found in the star-forming clumps in the tails of \acrlong{gasp} galaxies \citep{poggianti2019-g13}.

\begin{figure*}
    \includegraphics[width=0.95\textwidth]{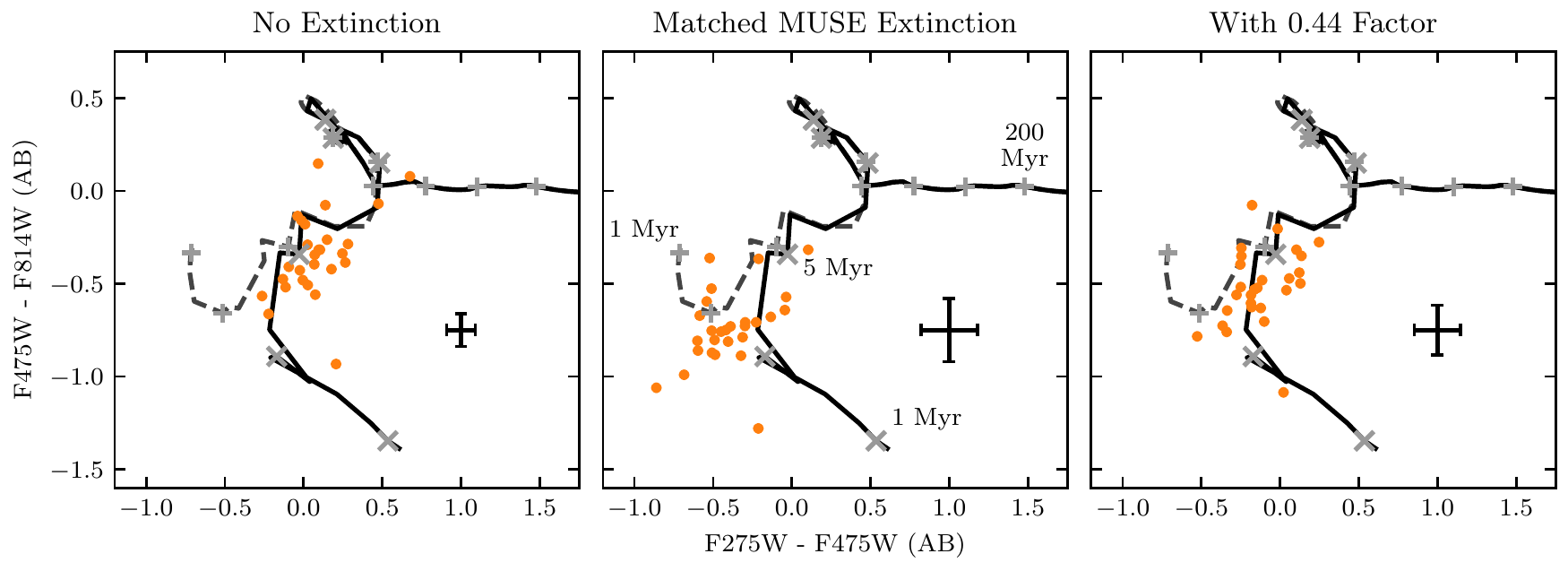}
    \vspace{-0.4cm}
    \caption{
    Colors of \gls{muse} \ion{H}{ii} regions on the Starburst99 track (dashed line) and
    the Starburst99 + Cloudy track (solid line) with different intrinsic extinction values.
    In each of the three plots, the scatter points in orange represent the combined photometry
    for all the \gls{hst} sources detected in each \ion{H}{ii} region.
    The median, minimum and maximum errors on each of these scatter points is
    0.14, 0.04 and 0.73 AB magnitude, respectively.
    Each plot also contains a Starburst99 track (gray, dashed line) created with the Genv00
    model \citep{ekstrom2011}, a \citet{kroupa2001} \gls{imf} and $Z = 0.014$.
    The ticks on each track are the same as those in Fig.~\ref{fig:colorComps}.
    Likewise, the Starburst99 + Cloudy track is presented as solid, black line.
    The big cross in each panel is the median uncertainty for the source colors,
    with different values of extinction adopted.
    \emph{Left:} No intrinsic extinction is applied on the colors of the \ion{H}{ii} regions.
    \emph{Middle:} The intrinsic extinction, determined from the \gls{muse} spectra of nebular
    emission, is applied on individual \ion{H}{ii} region.
    \emph{Right:} The same intrinsic extinction as used in the middle panel, but with a
    stellar-to-nebula extinction ratio of 0.44 \citep{calzetti1997}, is applied on individual
    \ion{H}{ii} regions. This choice of intrinsic extinction provides the best match between the
    \gls{hst} colors and the model.
    This plot also shows that the blue star clusters detected by \gls{hst} are indeed young
    with ages of~$<$~10\,Myr.
    }
    \label{fig:museRegsWithTrack}
\end{figure*}

With the intrinsic extinction of these \ion{H}{ii} regions constrained, we can compare
their \gls{hst} colors with the \gls{ssp} track, which is done in
Fig.~\ref{fig:museRegsWithTrack} as discussed in the following.
For each \ion{H}{ii} region, since colors of nearby \gls{hst} sources tend to be similar
(see Section~\ref{sub:hii} and Fig.~\ref{fig:clumpColors}), and considering
\gls{muse}'s much lower angular resolution than the \gls{hst} images, we combine all
\gls{hst} sources within the aperture to derive the colors of the total light.
The left panel of Fig.~\ref{fig:museRegsWithTrack} shows the \gls{hst} colors of the
\ion{H}{ii} regions, without the correction for the intrinsic extinction.
The comparison with the \gls{ssp} tracks shows, not surprisingly, that the observed
colors (especially F275W - F475W) are generally too red, as it is expected that most
of these \ion{H}{ii} regions are younger than $\sim$ 7 Myr.

\begin{figure}
    \centering
    \includegraphics{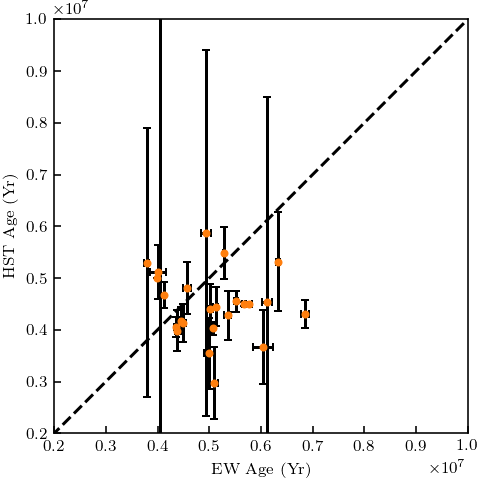}
    \vspace{-0.25cm}
    \caption{The \gls{ssp} age derived from the H$\alpha$ \gls{ew} vs. the \gls{ssp} age derived from the \gls{hst} broad-band colors as shown in Fig.~\ref{fig:colorComps}. Both sets of age estimates were predicted using the Starburst99 + Cloudy model.
    We limit the comparison to \ion{H}{ii} regions without bright stars within $1\farcs{}5$.
    The best-fit relation is close to the unity line that suggests a general good consistency between these two age estimates, given the current uncertainty.
    }
    \label{fig:museHstAgeComp}
\end{figure}

It should be noted that the Starburst99 \gls{ssp} models do not include nebular emission
from the warm and ionized gas, which can be
significant for young stellar populations (e.g., age $<$ 10 Myr).
We ran the development version of the photoionization code Cloudy, last reviewed by \citet{cloudy2017},
to add nebular emission to the stellar component of the radiation field reported by Starburst99.
Cloudy does a full ab initio simulation of the emitting plasma, and solves self-consistently
for the thermal and ionization balance of a cloud, while transferring the radiation through the cloud
to predict its emergent spectrum.
We assumed a nebula of density 100~cm$^{-3}$, and metallicity of~0.7 solar, surrounding the stellar source and
extending out to~1\,kpc from it. For the inner radius of the cloud, we experimented with two values~(1\,pc and~10\,pc), but found that the predicted colors do not depend on that choice.
We also imposed a lower limit of~1\% on the electron fraction to let the
calculation extend beyond the \ion{H}{II} region, into the photo-dissociation region. The Cloudy
modification is only important for star clusters younger than 10 Myr (Fig.~\ref{fig:museRegsWithTrack})
but does help to explain the F475W - F814W color for some sources.
The inclusion of the nebular lines and continuum affects the F475W flux the most and the F275W flux the least.

The middle panel of Fig.~\ref{fig:museRegsWithTrack} shows shows the colors of \ion{H}{ii} regions,
after the correction for the intrinsic extinction derived from the \gls{muse} data.
The same tracks as on the left panel, Starburst99 and Starburst99 + Cloudy, are also plotted.
While this comparison does suggest these \ion{H}{ii} regions are young (e.g., age $<$ 7 Myr),
the F275W - F475W colors are typically too blue.

One way to alleviate this discrepancy is to consider the extinction difference between stars and nebulae.
It has been known that the measured extinction on the stellar light can be different from the measured
extinction on the warm gas \citep[e.g.,][]{calzetti1994,calzetti1997,koyama2019}.
\citet{calzetti1997} gave an average relation of E(B-V)$_{\mathrm{star}}$ / E(B-V)$_{\mathrm{gas}}$ = 0.44
\citep[also see][]{calzetti1994,calzetti2000}.
This difference may suggest the spatial decoupling of the ionized gas and the young stellar population and other geometry effects
\citep[e.g.,][]{calzetti1994,charlot2000}.
There have been some works to study the relation between this ratio and specific star formation rates,
redshift and stellar mass \citep{wild2011,wuyts2011,price2014,reddy2015}.
Most recently, \citet{koyama2019} has shown that this ratio generally increases with increasing specific
star formation rate while it decreases with increasing stellar mass, although the scatter is substantial.
In this work, we simply apply the ratio of 0.44 as suggested by \citet{calzetti1997}.
With this factor included, as shown in the right panel of Fig.~\ref{fig:museRegsWithTrack},
the match between the \gls{hst} colors and the Starburst99 + Cloudy model is improved.

To summarize, we include two corrections, adding nebular emission from Cloudy and considering
the different extinction for stars and gas, to alleviate the initial discrepancy between the
\gls{hst} broad-band colors and the \gls{ssp} tracks.
Given the uncertainty on the \gls{hst} colors, the intrinsic extinction on young stars,
the Balmer decrement and the Starburst99 models, the colors of these \ion{H}{ii} regions are
consistent with the expectation for young stellar populations at age of~$<$~10\,Myr.
Therefore, in this work, we simply adopt an intrinsic extinction of
$E(B-V) = 0.08$
for all the \gls{hst} sources in the tail.
This value is derived from $0.44 A_\mathrm{V} = E(B-V) k_\mathrm{V}$ where
$A_\mathrm{V} =0.72$\,mag from the \gls{muse} data, $k_\mathrm{V}$ is calculated
according to the \citet{calzetti2000} law with $R_\mathrm{V} = 4.05$.
Fig.~\ref{fig:colorComps} then compares the colors of the \gls{hst} sources to these tracks.

We also compare the age determined from the \gls{hst} broad-band colors with the age determined
from the H$\alpha$ \gls{ew} especially since the \gls{ew} is not affected by the intrinsic extinction.
The Starburst99 model gives the direct relation between the \gls{ssp} age and the H$\alpha$ \gls{ew}.
Using the same model and the associated track in Fig.~\ref{fig:museRegsWithTrack}, we also predicted the age of sources using the \gls{hst} colors. The age was determined by matching the colors of sources in Fig.~\ref{fig:museRegsWithTrack}-\emph{right} to the track using the shortest euclidean distance. The age error is estimated from Monte Carlo simulations with the errors of colors considered.
As shown in Fig.~\ref{fig:museHstAgeComp}, the consistency between two age estimates is generally good, when we only include \ion{H}{ii} regions with robust \gls{ew} measurements.
The best-fit relation is also close to the unity line. Thus, with all the uncertainty discussed
above, we conclude that the \gls{hst} broad-band colors can present good constraints of the \gls{ssp} age.

%% file: sections/s4-youngclusters/subs/props.tex
\subsection{Properties of the HII regions}\label{sub:sfProps}

\begin{figure*}
    \includegraphics[width=0.9\textwidth]{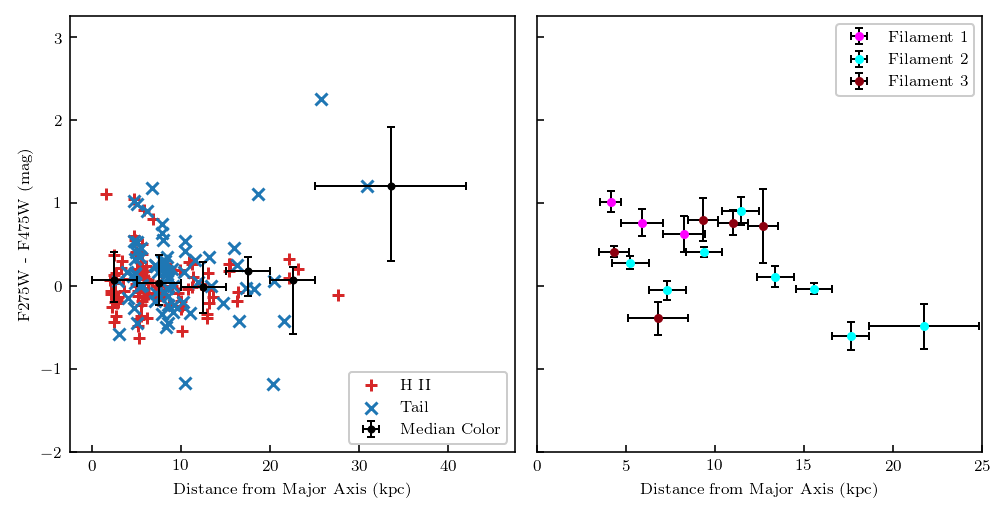}
    \vspace{-0.4cm}
    \caption{
            \emph{Left}:
            The F275W - F475W color vs. distance relation for a subset of sources in the \ion{H}{II} and tail samples
            established in Section~\ref{sub:colorResults}. We restrict the sources to have an error of less than 0.4~mag here rather than~1.0~mag as in Section~\ref{sub:colorResults}.
            The median colors plotted include both \ion{H}{II} and tail sources. The $x$
            error bars indicate the bin size and the $y$ error bars indicate the 1-$\sigma$ scatter in the bin.
            There is no clear change on the average source color with the distance to the galaxy.
            \emph{Right}:~%
            The same relation for three different filaments extending from \eso{}.
            Filament~1 corresponds to the filament with regions~25 and~26 in Fig.~\ref{fig:wholeWithRegs},
            Filament~2 corresponds to the filament with regions~23 and~27-30 in
            Fig.~\ref{fig:wholeWithRegs}, and Filament~3 corresponds to the filament with regions~17,~18 and~21 in Fig.~\ref{fig:wholeWithRegs}. Although the uncertainties are still large and the intrinsic extinction is not applied, at least two filaments become bluer further away from the galaxy.
            }
    \label{fig:colByDist}
\end{figure*}

We also analyzed the correlation between the source distance from the major axis of the
galaxy and the color of that source. Fig.~\ref{fig:colByDist} (left) shows that there
is no strong evidence for the change of the F275W - F475W color with the distance to
the galaxy, with the large scatter of the colors observed.

\begin{figure*}
    \includegraphics[width=0.75\textwidth]{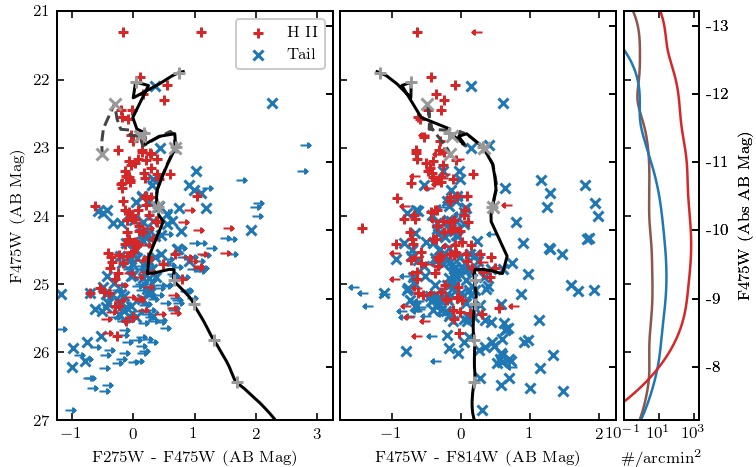}
    \vspace{-0.2cm}
    \caption{
        The color-magnitude diagram for the sources in the \ion{H}{ii} and tail regions, with the absolute magnitude also shown on
        the right side.
        The same tracks as in Fig.~\ref{fig:museRegsWithTrack} are also shown (for a total mass of $10^4$ M$_{\sun}$).
        The markers on tracks are also the same (note star clusters become dimmer as they get older so small ages start from the top).
        From these tracks, the mass of these young star complexes is
        $10^4$ - $10^5$ M$_{\sun}$ if younger than $\sim$100 Myr.
    }
    \label{fig:colorMagDiagram}
\end{figure*}

Fig.~\ref{fig:colByDist} (right) shows the color along three different blue filaments within the tail.
This approach was chosen to ensure no information was lost in the ensemble method shown in
Fig.~\ref{fig:colByDist} (left). All light within the filament regions (less the bright foreground
sources) is integrated for this second figure to see how the color changes along an individual filament.
Although it is not significant, there is a slight trend from red to blue for filament~1 as the distance increases. Filament~3
has an interesting discontinuity where the color goes from blue to red then to blue again as the distance
increases.

We also present the Color-Magnitude diagram in Fig.~\ref{fig:colorMagDiagram} which gives us
a constraint on the masses of the young star complexes if the age is determined from e.g., the color-color diagram.
The color-magnitude diagram suggests that the young star complexes have a mass of
$10^4$-$10^5$ M$_{\sun}$ if younger than $\sim$100 Myr
(see more detail regarding mass calculations in Section~\ref{sub:tailSFH}).
The diagram on the left also delineates the detection threshold
near the bottom, between the F475W magnitude and the F275W-F475W color.
The displayed track is the same as the ones used in
Fig.~\ref{fig:colorComps} and Fig.~\ref{fig:museRegsWithTrack}.

\begin{figure}
    \includegraphics[width=\columnwidth]{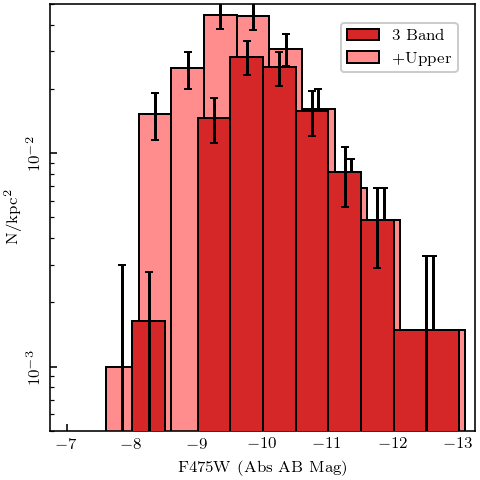}
    \vspace{-0.4cm}
    \caption{
        The luminosity distribution for sources detected in the \hii{} and tail regions defined in Section~\ref{sub:roi} with the exception that a
        F275W - F475W color cutoff of~2.5 is chosen. The control sample has been subtracted from
        each distribution. The darker distribution represents the sources detected in three bands
        whereas the lighter distribution include F275W upper limits. Note that the upper limit distribution has been offset for display only.
        }
    \label{fig:luminDistrib}
\end{figure}

The luminosity distribution is presented in Fig.~\ref{fig:luminDistrib}.
The figure shows \ion{H}{II} and tail sources (see Section~\ref{sub:roi}) that were detected in all three bands (red) and F475W/F814W detections + F275W upper limit (light red).
We model the data in Fig.~\ref{fig:luminDistrib} according to a luminosity function of $dN/dL \sim L^{-a}$ where
$dN$ is the number counted per luminosity bin $dL$, $L$ is the luminosity and $a$ is the power law index.
The three band detections (for sources brighter than -9.5 mag) correspond to $a = 2.0 \pm 0.1$ while the
addition of upper band detections correspond to $a = 2.3 \pm 0.1$.

%% file: sections/s4-youngclusters/subs/muse_alma.tex
\subsection{Relationship with the \ion{H}{II} regions and CO sources}
\label{sub:muse_alma}

To compare the distribution of the young star complexes, the \ion{H}{II} regions, and the CO compact sources, we need to verify the astrometry between \gls{hst}, \gls{muse} and \gls{alma}.
First, we compared the
astrometry of the \gls{hst} images and the \gls{muse}
continuum surface brightness map.
We select 16 bright
stars from the {\em 2MASS} point source catalog
in the common area of the \gls{hst} and \gls{muse} fields, which are adopted as the
reference stars of the WCS for each field. Then we use the WCSTools package to update the WCS of the \gls{hst} images
and \gls{muse} H$\alpha$ surface brightness map. By comparing the position of the reference stars in each image, we found that the offsets between
them are generally (75\%) less than the pixel size $0\farcs2$ of the \gls{muse} maps. The rms of $\delta$RA and $\delta$Dec
are $0\farcs22$ and $0\farcs05$, respectively, suggesting that the WCS of these two images are well aligned. This first step has already calibrated the absolute astrometry of the \hst{} and \gls{muse} data.
Second, we in principle need to compare the astrometry of the \gls{muse} H$\alpha$ surface
brightness map and the \gls{alma} CO(2$-$1) intensity map.
However, there is not a single common source between the \gls{muse} maps (continuum or lines)
and the \gls{alma} continuum map. One also cannot assume perfect matches between \gls{alma}
CO clumps and \gls{muse} \ion{H}{II} regions (see Fig.~\ref{fig:wholeWithRegs}).
Thus, we rely on the absolute astrometry of the \gls{alma} maps, which should be better than
$0\farcs3$ for our data (see the \gls{alma} website%
\footnote{https://help.almascience.org/kb/articles/what-is-the-absolute-astrometric-accuracy-of-alma}
on the typical absolute astrometric accuracy of \gls{alma}).

As shown in Fig.~\ref{fig:wholeWithRegs}, the correlation between the \hst{} blue star clusters and the \ion{H}{II} regions is very good, with all \gls{muse} \ion{H}{II} regions having at least one \hst{} blue star cluster within 0.2 kpc. There are many \hst{} blue sources not identified as \ion{H}{II} regions from our criteria but they are typically associated with faint H$\alpha$ clumps. On the other hand, only about a quarter of \ion{H}{II} regions have associated CO clumps within 0.3 kpc, with another quarter having nearby CO clumps beyond 0.3 kpc and half of \ion{H}{II} regions without nearby CO clumps detected at all. Some CO clumps are also not associated with any activity of \gls{sf}, being \ion{H}{II} regions or the \hst{} blue star clusters, e.g., in zoom-out \#11, \#16, \#22, \#34, \#35 and \#36. Assuming \ion{H}{II} regions and \hst{} blue star clusters are formed out of molecular clouds, the parent molecular clouds may get disrupted quickly after the initial \gls{sf}.
Some molecular clouds probably have not collapsed yet. More detailed studies on the relationship between molecular clouds and young star complexes require detailed studies with the \gls{alma}, \gls{muse} and \hst{} data, which is beyond the scope of this paper.

%% file: sections/s4-youngclusters/subs/chandra.tex
\subsection{Relationship with the \chandra{} X-ray sources}
\label{sub:chandraComp}

\citetalias{sun2010} discovered some X-ray point sources around \eso{}'s tail with the \chandra{} data. Some of them were suspected to be associated with young star complexes in the tail. We zoom in around some \chandra{} sources close
to blue star clusters in Fig.~\ref{fig:wholeWithRegs}. Particularly, sources C6, C8, C9, C10, C11 and C12
are close to young star complexes. All of them are downstream and still close to the galaxy. If associated with
\eso{}, they would be \glspl{ulx}. The offset between the X-ray source and the young star complexes observed here, typically within several hundred pc, is also normal for \glspl{ulx} \citep[e.g.,][]{Poutanen13}.

%% file: sections/s6-discuss/discuss.tex
\section{Discussion}\label{sec:discuss}

\input{sections/s6-discuss/subs/galSFH}

\input{sections/s6-discuss/subs/tailSFH}

\input{sections/s6-discuss/subs/strip_history}

%% file: sections/s6-discuss/subs/galSFH.tex
\subsection{Star formation in the galaxy}\label{sub:galSFH}

The total FIR luminosity of the galaxy was derived from the \emph{Herschel} data. We used the
\emph{Herschel} source catalog, particularly the PACS Point Source Catalog and the SPIRE Point
Source Catalog.
We used the python code MBB\_EMCEE to fit modified blackbodies to photometry data
using an affine invariant \gls{mcmc} method \citep{foreman2013}, with the \emph{Herschel}
passband response folded \citep{dowell2014}.
Assuming that all dust grains share a single temperature $T_\mathrm{d}$, that the dust distribution
is optically thin in the FIR, and neglecting any power-law component towards shorter wavelengths, the fit
results in a temperature of $T_\mathrm{d}$ = (32.1$\pm$0.5) K, a luminosity
$L_{\mathrm{8-1000 \ \mu m}}$ = (5.24 $\pm$ 0.12) $\times$ 10$^{9}$ L$_{\odot}$, and a dust mass of
$M_\mathrm{d}$ = (1.3 $\pm$ 0.1) $\times 10^{6}$ M$_{\odot}$ for $\beta$ = 1.5.
For $\beta$ = 2, $T_\mathrm{d}$ = (28.6$\pm$0.5) K,
$L_{\mathrm{8-1000 \ \mu m}}$ = (5.06 $\pm$ 0.11) $\times$ 10$^{9}$ L$_{\odot}$ and
$M_\mathrm{d}$ = (2.1 $\pm$ 0.2) $\times 10^{6}$ M$_{\odot}$.
It is noted that the dust temperature in \eso{} is higher than those typically found in Virgo
cluster galaxies \citep[$\sim$ 20 K,][]{davies2012, auld2013}.
Such a high $T_\mathrm{d}$ is consistent with the results by \citet{boccio2014}.

The total \gls{sfr} of the galaxy is 0.97 M$_{\odot}$/yr, from the \emph{Galex} NUV flux density and the
total \emph{Herschel} FIR luminosity with the relation from \citet{hao2011}. If using the
\emph{WISE} 22 $\mu$m flux density and the relation from \citet{lee2013}, the estimated total \gls{sfr} is
1.39 M$_{\odot}$/yr. The Kroupa \acrfull{imf} is assumed in both cases. The \citet{lee2013} work assumed the
Salpeter \gls{imf} so we multiply its \gls{sfr} relation by 0.62 to convert to the Kroupa \gls{imf}.
With the measured molecular gas content of $\sim 1.1\times10^{9}$ M$_{\odot}$ in the galaxy \citep{jachym2014}, the gas depletion timescale is $\sim$ 0.9 Gyr.
\eso{} still appears on the galaxy main sequence with its current \gls{sfr} and the total stellar mass \citep{Boselli2022}. 

The upstream of the galaxy (or the east side, or the near side) is dust free and gas free \citep{sun2007HA,sun2010,fossati2016} so the
current \gls{sf} is mainly around the nucleus and the downstream. The H$\alpha$ disk is truncated to $\sim$ 1.5 kpc radius, similar to the size of the remaining CO cloud in the galaxy (Fig.~\ref{fig:esoImageAllFilt}).
More detailed studies of SF history in the galaxy can be done with the {\em MUSE} data and multi-band photometry data in the future.

%% file: sections/s6-discuss/subs/tailSFH.tex
\subsection{Star formation in the tail}\label{sub:tailSFH}

\begin{figure}
    \centering
    \includegraphics[width=0.9\columnwidth]{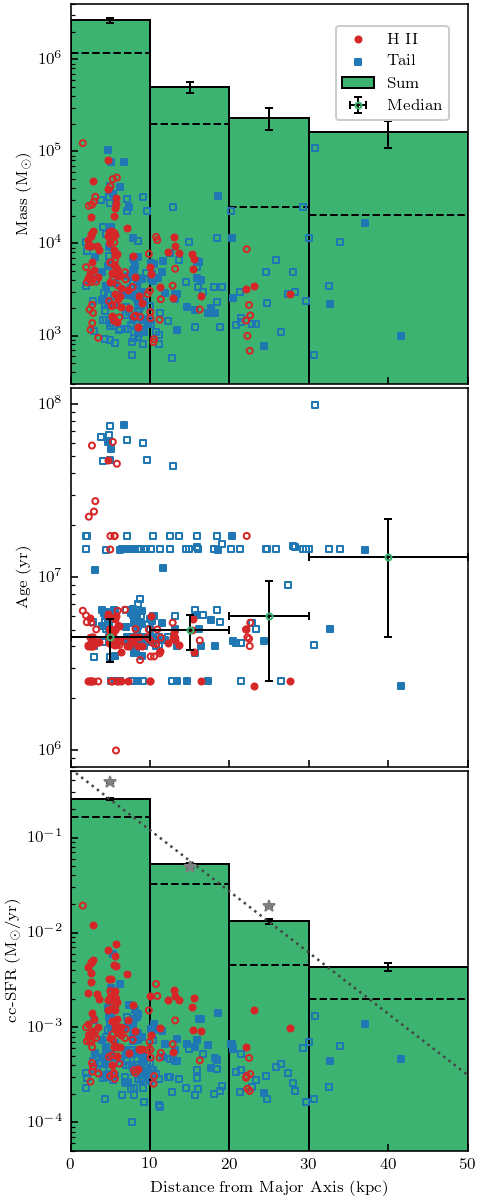}
    \vspace{-0.2cm}
    \caption{
    The mass, age, and \acrfull{esfr} for sources younger than~100\,Myr as estimated with the
    Starburst99 + Cloudy model. For the age distribution, the median values (filled
    markers for three-band detections and empty markers for two-band detections) and~1$\sigma$
    scatter are also shown.
    The histograms in the top and bottom panels represent the median summed masses and
    \glspl{esfr} of both regions within the respective distance bin, with the uncertainty also shown.
    The dashed histograms and filled green scatter points are for the same quantities but
    for the~3 band detections only.
    The dotted line in the bottom panel shows an exponential fit of the total \gls{esfr} to the source distance (see Section~\ref{sub:tailSFH}).
    The large grey stars in the bottom panel show the total \gls{sfr} derived from the H$\alpha$ data in the same bin.
    }
    \label{fig:sfrVsDist}
\end{figure}

\Acrlong{sf} in the tail can be constrained from the H$\alpha$ data.
As discussed in Section~\ref{sub:roi}, we defined~43 \ion{H}{ii} regions in the tail region,
including~37 beyond the galaxy region.
The total H$\alpha$ flux of each region, within a circular aperture with a radius of~1.4$''$,
is measured, after correcting for both the Galactic extinction and the intrinsic extinction.
The total H$\alpha$ luminosity for~43 \ion{H}{ii} regions is~8.1$\times10^{40}$\,erg\,s$^{-1}$.
Excluding the six in the galaxy region, the total H$\alpha$ luminosity
is~4.0$\times10^{40}$\,erg\,s$^{-1}$.
With the H$\alpha$ --- \gls{sfr} relation from \cite{hao2011} assuming a Kroupa \gls{imf},
the corresponding \gls{sfr} is~0.45 and~0.22\,M$_{\sun}$/yr, respectively.
These \gls{sfr} values are similar to the estimate from
\citetalias{sun2007HA}~(0.59 M$_{\sun}$/yr) for~29 \ion{H}{ii} regions assuming a
Salpeter \gls{imf} and $A_{\rm V}$ = 1 mag.

\Acrlong{sf} in the tail can also be constrained from the \gls{hst} broad-band photometry
data as discussed in Section~\ref{sec:youngClusters}. 
We used the Starburst99 + Cloudy model in Fig.~\ref{fig:colorComps} to estimate
the ages and masses of star clusters.
The age was estimated by comparing the F275W-F475W and F475W-F814W colors to the
track in Fig.~\ref{fig:colorComps}, taking the value at the closest distance from the track.
This process was performed with a Monte Carlo simulation using 50,000 samples where the
F275W, F475W, and F814W magnitudes were assumed to be normally distributed using the measured
magnitude as the mean and the measured uncertainty as the standard deviation.
Again, the assumed intrinsic extinction is $E(B-V)=0.08$ as discussed before.
Once an age is determined, we estimate the mass of the source by matching the F475W magnitude
to the corresponding track in the color - magnitude relation in Fig.~\ref{fig:colorMagDiagram}.
This is done by assuming the difference in measured F475W magnitude and track magnitude for the matched
age is due to the difference in mass.
For example, the Starburst99 track in Fig.~\ref{fig:colorComps} assumes a population mass of~$10^6$\,M$_{\sun{}}$.
If a source is 5\,mag dimmer in F475W than the age-matched point on the Starburst99 track (after the correction on the intrinsic extinction), the
mass is then $10^4$\,M$_{\sun{}}$.
Once the age and mass are known, we can estimate the \acrfull{esfr} by
dividing the mass by the age and sum up
the individual \gls{esfr} values for all sources to get the total \gls{esfr} in each Monte Carlo iteration.%
\footnote{We add ``color-color'' ahead of \gls{sfr} to distinguish from \gls{sfr} values typically estimated from monochromatic luminosity (e.g., H$\alpha$) or combination (e.g., FIR + UV), which is calibrated assuming continuous \gls{sf} over the timescale probed by the specific emission being used \citep[e.g.,][]{hao2011}.
}
If only sources younger than~10\,Myr are counted, the median \gls{esfr} is 0.199\,M$_{\sun}$/yr and~0.083\,M$_{\sun}$/yr for the \ion{H}{II}
and tail sources, respectively.
If instead all sources younger than~100\,Myr are included, the \gls{esfr} is 0.210\,M$_{\sun}$/yr and~0.117\,M$_{\sun}$/yr for the \ion{H}{II}
and tail sources, respectively. The Monte Carlo uncertainty on these values is~0.7\% and~6.4\% for the \ion{H}{II}
and tail sources, respectively, according to the \acrlong{mad}.
This estimate of the \gls{esfr} for \ion{H}{II} regions is only about half the estimated \gls{sfr} from the H$\alpha$ data (see Section~\ref{app:sfrComp} and Fig.~\ref{fig:sfrComp}).

The total \gls{esfr} in Fig.~\ref{fig:sfrVsDist} was fit to the source distance ($d$)
(dark-gray dotted line in bottom panel) according to
\begin{equation}\label{eq:sfrModel}
    \mathrm{cc\mbox{--}SFR} = a \exp \left( -d/\delta \right).
\end{equation}
The resulting fit gives $a = 0.53\pm0.01$\,M$_{\sun{}}$\,yr$^{-1}$ and $\delta = 6.7\pm0.4$\,kpc, which again shows the \gls{esfr} in the tail decreases fast with the distance to the galaxy.
We also derived the \gls{sfr} from the H$\alpha$ data in the same spatial bins (Fig.~\ref{fig:sfrVsDist}). The fit with the same model gives $a = 0.89\pm0.04$\,M$_{\sun{}}$\,yr$^{-1}$ and $\delta = 5.7\pm0.6$\,kpc.
\citet{cramer2019} noted this trend in D100. This trend is also noted for some other galaxies in \gls{rps} \citep[e.g.,][]{george2018,poggianti2019-g13,Boselli2022}.

With E(B-V)$_{\mathrm{star}}$ / E(B-V)$_{\mathrm{gas}}$ = 0.44, the total stellar mass in
the tail is $\sim 2.7 \times 10^6$\,M$_{\sun{}}$. With the above ratio equal to 1, the total stellar
mass remains about the same, $\sim 2.9 \times 10^6$\,M$_{\sun{}}$.
Fig.~\ref{fig:sfrVsDist} shows the Monte Carlo median mass, age, and \gls{esfr} for
sources less than~100\,Myr old as a function of distance from the galactic major axis for the
three band and two band young star clusters identified in Section~\ref{sub:roi}.
Likewise, the graphic shows the median summed masses, median ages, and median summed \glspl{esfr}
for four distance bins.
The individual sources do not show any significant trend regarding how these
properties change with distance.
However, the summed and median statistics do show that the total mass and the total
\gls{esfr} decrease with distance from the galaxy while the median age of the sources remains nearly constant with the uncertainty.
On the other hand, any age trend is limited by our sensitivity to old~($>$~100\,Myr) sources as discussed later in this section.
About~95\% of \gls{sf} in the tail is within~20\,kpc from the galaxy.
The \gls{sf} beyond~20\,kpc is still observed but very weak.

Fig.~\ref{fig:sfrVsDist} shows that sources older than~40\,Myr are mostly close to the galaxy (within~13\,kpc). There is a lack of old sources beyond~15\,kpc from the galaxy. There are several factors that can contribute to this result. First, assuming a galaxy total mass of $5 \times 10^{10}$\,M$_{\sun{}}$, the free fall time from~30\,kpc to the galactic plane is $\sim$ 350\,Myr. Considering, the actual fall-back time to any height above the galactic plane would be some fraction of this value, it is
conceivable that some old sources that are close to the galaxy~($<$~10\,kpc)
formed further away from the galaxy, but have had sufficient time to fall back toward the galaxy.
Second, the intrinsic extinction may be underestimated for some old sources that are close to the galaxy, which would result in over-estimate of their ages.
Third, the faint, old (or red) sources have a larger contamination from background sources (see the \glspl{kde}
in Fig.~\ref{fig:colorComps}).

\begin{figure}
    \centering
    \includegraphics[width=0.9\columnwidth]{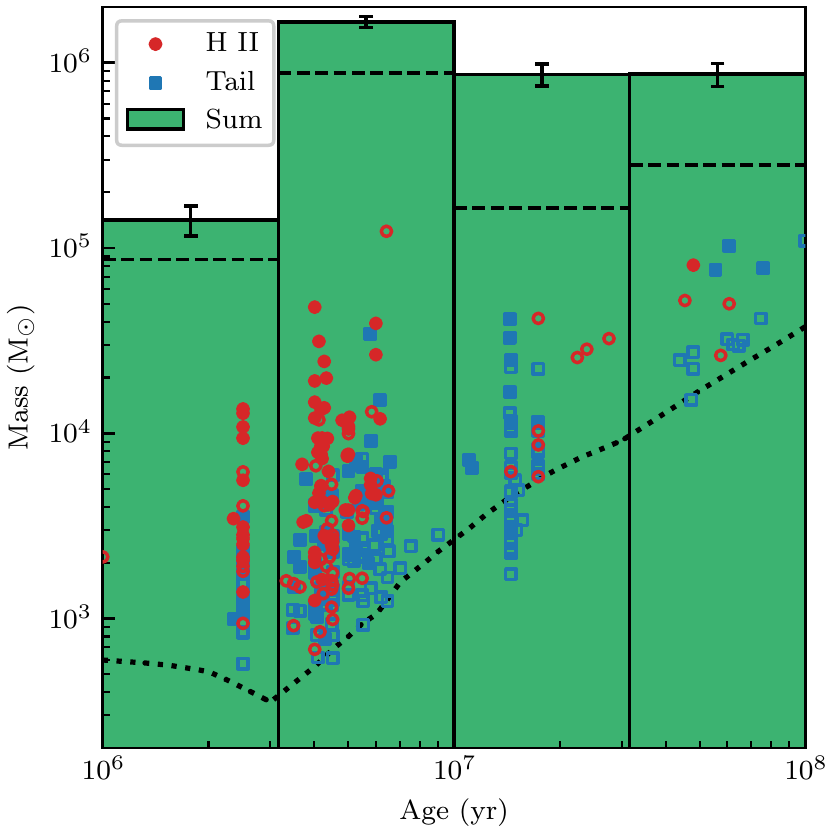}
    \vspace{-0.2cm}
    \caption{
    The mass
    with the source age for sources younger than~100\,Myr as
    estimated with the Starburst99 + Cloudy model.
    The marker and histogram styles are the same as in Fig.~\ref{fig:sfrVsDist}. The empirical mass limit that can be reached with our data is shown as the dotted line.
    }
    \label{fig:massSfrVsAge}
\end{figure}

Fig.~\ref{fig:massSfrVsAge} shows the estimated mass
for sources younger than~100\,Myr as a function of the source age for the three band and two band young star complexes identified in Section~\ref{sub:roi}.
On the face value, Fig.~\ref{fig:massSfrVsAge} suggests that most \gls{sf} in the tail happened within the recent 3 - 10\,Myr. However, as sources become fainter when they age, it is clear that the current data present a lower mass detection limit increasing fast with the source age (also evident in Fig.~\ref{fig:massSfrVsAge}).
We then derived the empirical mass limit in Fig.~\ref{fig:massSfrVsAge} to better understand this limitation.
The displayed limit is calculated by requiring the error of the F275W - F475W color less than 1 and also assuming the Starburst99 + Cloudy track in Fig.~\ref{fig:colorComps} and the median extinction
reported in Fig.~\ref{fig:museEWvsAV}.
As shown in Fig.~\ref{fig:massSfrVsAge}, the displayed limit does well to predict the limiting
mass at each age helping explain why we do not measure high age / low mass sources.
A few sources do exist below this empirical limit, however they tend to be sources detected in
two bands (and are therefore likely to not be detected in F275W at all).
We fit the mass limit for ages greater than~3\,Myr according to a power law with
an index of~1.3.

Fig.~\ref{fig:massSfrVsAge} also reveals a population of young~($<$~10\,Myr) tail sources that are not associated the an \ion{H}{II} region. As shown in Fig.~\ref{fig:colorComps}, the background contamination for these blue sources is negligible. The total mass of these sources sums to
$3 \times 10^5$\,M$_{\sun{}}$.
However, some of these sources are near \ion{H}{II} regions (see Fig.~\ref{fig:srcPositions})
although they are not in the \ion{H}{II} regions of interest detailed in Section~\ref{sub:roi}.
We therefore measure the mass of these tail sources that are within~0.7\,kpc of \ion{H}{II} regions to be
$1.5 \times 10^5$\,M$_{\sun{}}$.
For the remaining sources, they are typically within~1.2\,kpc of the selected \ion{H}{II} regions, especially for bright or massive sources.

\begin{figure}
    \centering
    \includegraphics[width=\columnwidth]{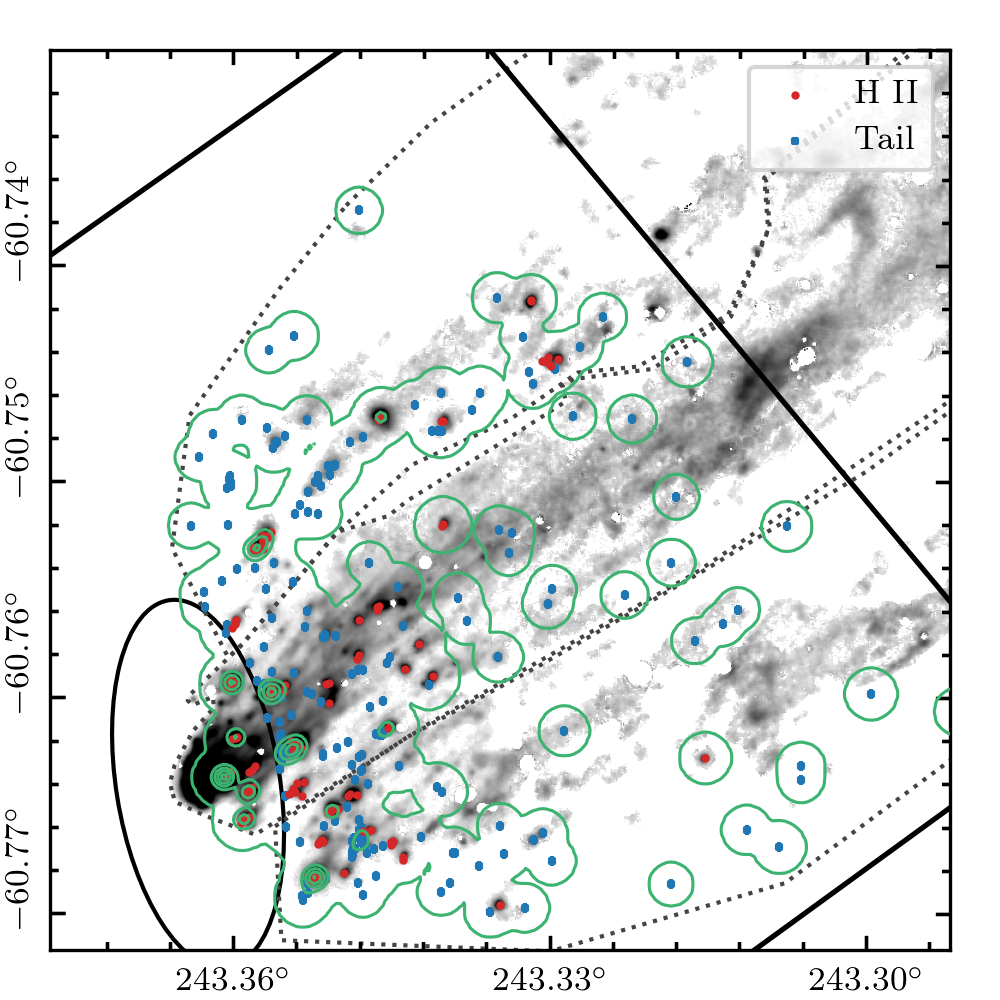}
    \caption{
    The \gls{esfr} distribution in the tail of \eso{} plotted on the \gls{muse} H$\alpha$ image.
    The green contours are generated using a spatial \gls{kde} using the \gls{esfr}
    of each source as the sample weight.
    Areas with higher contour value correspond to 1) a higher density of young star clusters
    and/or 2) regions of high \gls{esfr}.
    The figure highlights a few pockets of high \gls{sf} (primarily close to the galaxy center)
    while the majority of sources only make a small contribution to the total \gls{sf}.
    The solid, black lines show the F275W \gls{fov},
    and the dark-gray dotted lines show the division of the north, central, and southern tail zones from top to bottom. The galaxy region defined in Fig.~\ref{fig:compRegions} is shown as the black ellipse.
    }
    \label{fig:sfrDistrib}
\end{figure}

We also present the \gls{esfr} spatial distribution in Fig.~\ref{fig:sfrDistrib}.
The figure highlights a few pockets of higher \gls{sf} (primarily close to the galaxy center)
while the majority of sources only make a small contribution to the total \gls{sf}.
Although not shown in Fig.~\ref{fig:sfrDistrib}, the mass distribution is nearly identical
to the displayed \gls{esfr} distribution.

Fig.~\ref{fig:sfrDistrib} also marks the three zones in the tail defined by the H$\alpha$ data from \gls{muse}, north, central and south zones (more detail in \citealt{Luo22}). We performed the same analysis for Fig.~\ref{fig:sfrVsDist} for each of the tail zones. 
This further analysis was motivated by the different morphology in each of the three zones.
In the north zone, stripping is in an advanced stage as the galaxy region is nearly cleared. \ion{H}{II} regions are detected to the largest distance to the galaxy in this zone.
In the south zone, stripping may be in a little less advanced stage than in the north zone, as the south side of the galaxy is more on the leading size to the \gls{icm} wind while the north side of the galaxy is more on the trailing side. The majority of \ion{H}{II} regions in the southern zone are within~$\sim$~10\,kpc of the galaxy major axis.
The remaining sources in this zone tend to have the same mass distribution as the rest of the
zone but tend to have a higher age distribution.
For the central zone, stripping is still ongoing around the nuclear region from the X-ray, H$\alpha$ and CO data. All \ion{H}{II} regions are within~$\sim$~15\,kpc of the major axis.
There is a lower density of non-\ion{H}{II} sources within this zone that tend to have the
same age as the rest of the sources in this zone, though they do tend to be low-mass.
In the end, the results in these separated zones are similar to those in Fig.~\ref{fig:sfrVsDist}.

One should also be aware of caveats in our analysis in this section.
First, our results are only based on the data from three \hst{} photometry bands. Model fitting was done from the color-color diagram, while a more rigorous statistical approach with more photometry bands \citep[e.g.,][]{Linden21,Moeller22} cannot be performed.
Second, the SSP models have their uncertainty (see the recent comparison by \citealt{Wofford16}).
Third, our results are based on the assumption that each star cluster complex we studied can be reasonably approximated by an instantaneous
burst (single) population.
Fourth, we only assumed a fixed, average intrinsic extinction.
Lastly,
when the total mass is less than $\sim$ 3000~M$_{\odot}$, the stellar IMF is not fully sampled so the results below this mass limit will have a larger uncertainty. While the more robust stellar population synthesis codes for such low-mass clusters exist (e.g., SLUG from \citealt{slug12}), many of star clusters in our work are more massive than 3000~M$_{\odot}$ and a more in-depth stochastical modeling is beyond the scope of this work.

%% file: sections/s6-discuss/subs/strip_history.tex
\subsection{Stripping history of ESO 137-001}\label{sub:striphistory}

\begin{figure*}
    \centering
    \hspace{-0.3cm}
    \includegraphics[width=0.85\textwidth]{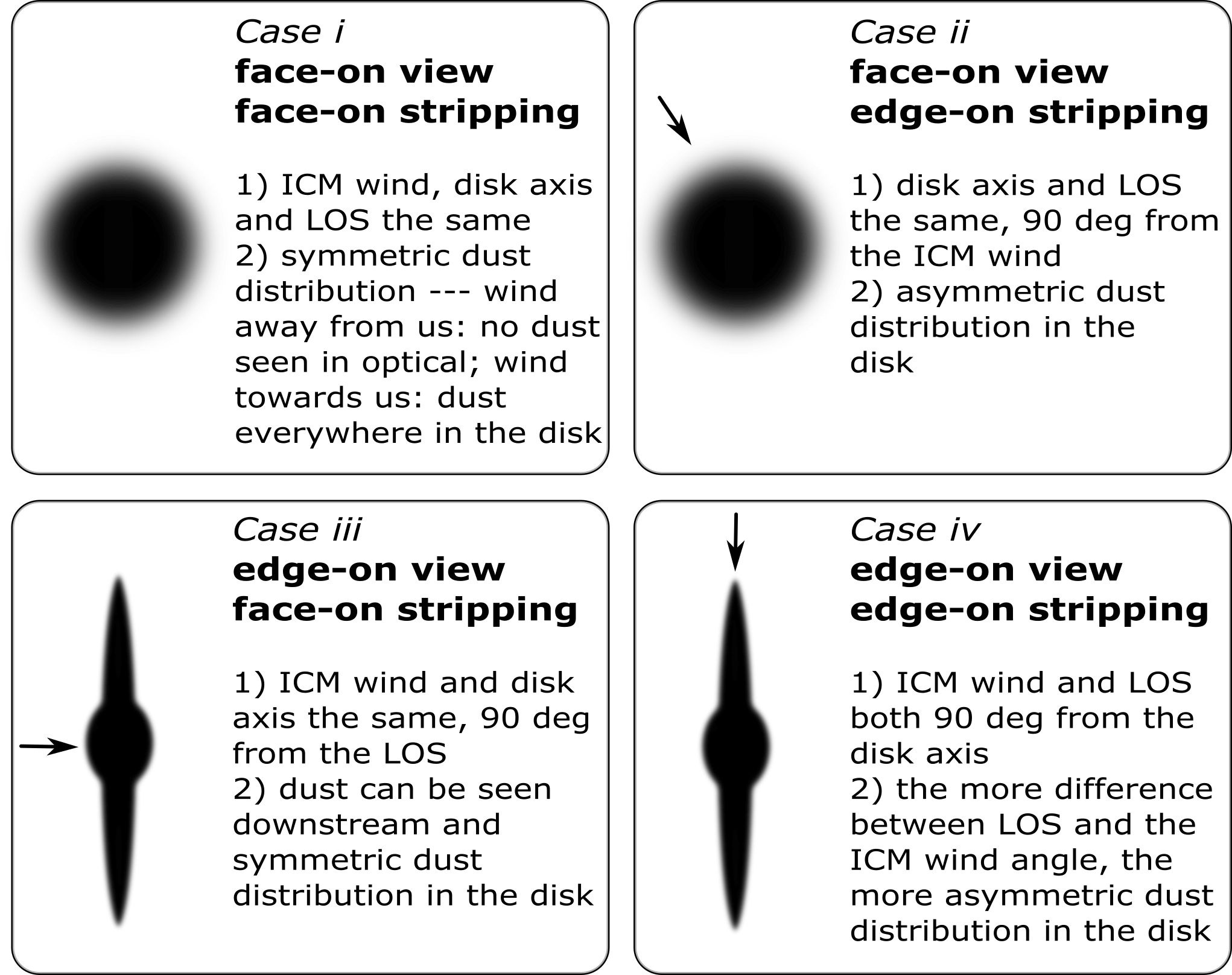}
    \vspace{-0.1cm}
    \caption{Four ideal examples of a symmetrical disk galaxy undergoing ram pressure stripping and the observed signature of dust distribution. The LOS is also perpendicular to the paper while the arrow marks the \gls{icm} wind direction.}
    \label{fig:galaxy_stripping}
\end{figure*}

What is the geometry of stripping in \eso{}?
The \hst{} data give the distribution of dust that can constrain the geometry of stripping.
Four special examples of a disk galaxy undergoing \gls{rps} and the expected signature of
dust distribution are shown in Fig.~\ref{fig:galaxy_stripping}. ESO~137-002 is close to case iv \citep{laudari2022} and NGC~4921 is close to case ii \citep{kenney2015}.
\eso{} has an inclination of $\sim 66^\circ{}$ so it is viewed closer to edge-on than face-on.
The eastern side of the galaxy is the leading side to ram pressure and is also the near side to us as discussed.
If \eso{} is moving on the plane of sky, the \gls{icm} wind angle with the disk plane
is $\sim 66^\circ{}$. As \eso{} has a
small velocity component towards us (relative to the cluster system velocity),
the \gls{icm} wind angle with the disk plane is less than $\sim 66^\circ{}$. Based on the tail direction, \cite{jachym2019} estimated an \gls{icm} wind angle of $\sim 47^\circ{}$, which makes stripping in \eso{} about the midway between edge-on and face-on.

What do the \hst{} data inform us on the stripping history of \eso{}?
As discussed in Section~\ref{sec:eso}, dust is detected around the nucleus and the
downstream region immediately behind the nuclear region.
\gls{rps} started from outside of the galaxy and has now progressed into the
central region of the galaxy.
\gls{sf} is still ongoing around the nucleus and the downstream region behind the nucleus,
as there is still abundant cold molecular gas around the nucleus (Fig.~\ref{fig:esoImageAllFilt}).
On the other hand, an ideal outside-in model of stripping is too simple.
One has to consider the actual distribution of the multi-phase \gls{ism} that is often porous so stripping can happen at multiple radii at the same time. The presence of a galactic bar may also cause a relative deficit of gas at intermediate radii.
As shown in Fig.~\ref{fig:esoProfile}, the colors along the major axis of the galaxy are mostly constant, while an ideal outside-in stripping and quenching may produce a continuous color gradient, as observed in D100 \citep{cramer2019}. More detailed analysis on the stripping/quenching in \eso{} is required in the future with the optical spectroscopic data from e.g., \gls{muse}.

A detailed inventory study of the \gls{ism} in \eso{} was done by \cite{jachym2014}.
About 80\% - 90\% of the original \gls{ism} in \eso{} has been removed from the galaxy,
presumably by \gls{rps}. At least half of the removed \gls{ism} is accounted for in the tail,
mostly in the molecular gas.
While more data are required to better constrain the mass the multi-phase gas in
\eso{}'s tail (e.g., \hi{} and improved H$\alpha$ estimates from the \gls{muse} data),
the biggest uncertainty seems to be on the diffuse cold molecular gas as the
\gls{alma} 12m + ACA data on CO(2-1) still miss $\sim$ 70\% of CO flux from the
single-dish data by {\em APEX} \citep{jachym2019}.

%% file: sections/s7-conc/conclude.tex
\section{Conclusions}\label{sec:conclude}

We present a detailed analysis of \eso{}, an archetypal \gls{rps} galaxy, with the \gls{hst} \gls{acs} and \gls{wfc3} data in four filters (F275W, F475W, F814W and F160W).
\begin{enumerate}

\item The galaxy has clear, asymmetric light and dust distribution indicative of ongoing \gls{rps} (Fig.~\ref{fig:wholeWithRegs}, Fig.~\ref{fig:esoImageAllFilt} and Fig.~\ref{fig:esoProfile}). 
The eastern side of the galaxy is the near side to us, also the leading side to ram pressure. The stripping is about the midway between edge-on and face-on stripping. The light profile effectively shows that \gls{sf} has been quenched in the upstream regions and the current \gls{sf} is mainly around the nucleus and downstream regions. The dust images show stripping near the nucleus of the galaxy where the dust has been pushed to the downstream side of the galaxy. We derived the E(B-V) map (Fig.~\ref{fig:extinction}) that shows the strong dust extinction downstream. There is also an enhanced dust feature at $\sim$ 2.3 kpc downstream, corresponding to a large CO clump. We suggest it is around the ``deadwater'' region. Stripping happens outside-in generally and has progressed into the inner $\sim$ 1.5 kpc radius of the nucleus. There is no evidence for an \gls{agn} in \eso{} from the \hst{} and X-ray data.
    
\item \hst{} data reveal active \gls{sf} in the downstream gas stripped (Fig.~\ref{fig:wholeWithRegs}, Fig.~\ref{fig:srcPositions} and Fig.~\ref{fig:colorComps}). We derived the color-color (F275W - F475W vs. F475W - F814W) diagram for sources identified in different regions of interest, including the galaxy, \hii{}, the tail and the control regions (Fig.~\ref{fig:compRegions}). The galaxy, tail and \hii{} regions all show significant excess of blue sources compared with the control region. We conclude these blue sources are young star complexes formed in the stripped \gls{ism} and \hst{} can pick up faint young star complexes no longer hosting bright \hii{} regions.
    
\item \hii{} regions in the stripped gas are well correlated with young, blue star clusters but not with CO clumps.
As shown in Fig.~\ref{fig:wholeWithRegs}, the correlation between the \hst{} blue star clusters and the \hii{} regions is very good, with all \gls{muse} \hii{} regions having at least one \hst{} blue star cluster within 0.2 kpc. Other \hst{} blue sources typically have faint H$\alpha$ clumps associated.
On the other hand, only about a quarter of \hii{} regions have associated CO clumps within 0.3 kpc, while half of \hii{} regions do not have nearby CO clumps detected at all. Some CO clumps are also not associated with any activity of \gls{sf}. We conclude that the parent molecular clouds get disrupted quickly after the initial \gls{sf}. Some molecular clouds are not forming stars at the moment. The comparison between the \hst{} and the {\em Chandra} images also suggests up to six \glspl{ulx} in the tail region.
    
\item Ages derived for the H$\alpha$ \gls{ew} are consistent with those derived from the \hst{} broadband colors (Fig.~\ref{fig:museRegsWithTrack} and Fig.~\ref{fig:museHstAgeComp}).
We applied a \gls{ssp} model with Starburst99 on these blue star clusters. For those associated with \hii{} regions with the \gls{muse} data, we can compare the age derived from the \gls{ssp} model (or with broadband colors) with the age derived from the H$\alpha$ \gls{ew}. While the initial analysis shows a significant discrepancy between two estimates, we conclude that these two estimates can be brought back into agreement if 1) allowing different extinction between the nebular and stellar components for young star complexes around \hii{} regions (particularly we adopted E(B-V)$_{\mathrm{star}}$ / E(B-V)$_{\mathrm{gas}}$ = 0.44 from previous studies); and 2) nebular emission is included in the \gls{ssp} tracks for ages of less than 20 Myr.
    
\item The \gls{sf} history in the tail can be quantitatively constrained from the \hst{} broad band colors.
We trace SF over at least 100 Myr and observe a broader spatial distribution of young star clusters than \hii{} regions only traced by H$\alpha$, and give a full picture of the recent evolutionary history of SF in the tail.
We showed that the average mass of the sources detected have a mass of $10^3$-$10^4$\,M$_{\sun{}}$ and that the ages of most sources is younger than approximately~100\,Myr. We measure the total \gls{sfr} of the \ion{H}{II} regions to be~0.2 - 0.45\,M$_{\sun{}}$/yr and other blue sources in the tail region add about 30\% more \gls{sfr}, all for sources younger than~100\,Myr. 
The total \gls{sfr} in the tail is substantial, about 40\% in the galaxy. We measure the total stellar mass in the tail to be $\sim 2.7 \times 10^6$\,M$_{\sun{}}$. The \ion{H}{II} and tail regions combined have a luminosity function $(dN/dL \sim L^{-a})$ for $a \approx 2.1$ (Fig.~\ref{fig:luminDistrib}).

\item We also examined the F275W - F475W color of selected sources in the \hii{} and tail regions as a function of distance from the galaxy (Fig.~\ref{fig:colByDist}) but no trend is found. The trend is also not clear for color changes along blue streams. While naively it is conceivable that the gas furthest from the galaxy was pushed out before the gas near the galaxy, the ages of young star complexes does not indicate this trend. Possible explanations include the distribution of the delay time between stripping and \gls{sf}, different \gls{sf} history for different star clusters.
    
\end{enumerate}

Our work demonstrates the importance of the \gls*{hst} data on the studies of \gls*{rps} galaxies.
More analysis with the data from \hst{} and \emph{James Webb Space Telescope}, and future wide-field survey data from {\em Euclid} and {\em Nancy Grace Roman Space Telescope} will allow us to better understand the young stellar population and SF efficiency in the RPS tails.

%% file: sections/acknowledge.tex
\section*{Acknowledgements}

We thank useful discussion with Hugh Crowl, Claus Leitherer and Renbin Yan.
We thank Nivedita Sekhar for some early work of the \hst{} data.
We thank the anonymous referee for useful comments.
Support for this work was provided by the National Aeronautics and Space Administration through \chandra{} Award Number GO2-13102A and GO6-17111X issued by the \chandra{} X-ray Center, which is operated by the Smithsonian Astrophysical Observatory for and on behalf of the National Aeronautics Space Administration under contract NAS8-03060. Support for this work was also provided by the NASA grants HST-GO-11683, HST-GO-12372.09, HST-GO-12756.08-A, GO4-15115X , NNX15AK29A and the NSF grant 1714764. P.J. acknowledges support from the project RVO:67985815, and the project LM2023059 of the Ministry of Education, Youth and Sports of the Czech Republic.

\section*{DATA AVAILABILITY}

The \hst{} raw data used in this paper are available to download at the The Barbara A. Mikulski Archive for Space Telescopes\footnote{\url{https://archive.stsci.edu/hst/}}. The {\em MUSE} raw data are available to download at the ESO Science Archive Facility\footnote{\url{http://archive.eso.org/cms.html}}. The reduced data underlying this paper will be shared on reasonable requests to the corresponding author.

%% file: sections/a1-phot/phot.tex
\section{Photometry}\label{app:phot}

\input{sections/a1-phot/subs/hst_phot_val}

%% file: sections/a1-phot/subs/hst_phot_val.tex
\subsection{\hst{} Photometry and Validation}
\label{sub:hstPhot}

We measured the photometry in each band with SExtractor~\citep{bertin1996} in dual image mode to ensure one-to-one mapping between detection and analysis sources. We tested this method using each of the four bands as the detection band and eventually chose to use F475W for detection as the F475W data are the most sensitive among all four bands to detect faint young star complexes. Our SExtractor setup was checked against the \citet{hammer2010} results where we were able to reproduce their results with our setup on the same data.
This work utilized 0\farcs{}5 apertures with corrections according to the \gls{hst}
encircled energy tables.

The analysis in this work was done with a combination of our own
software\footnote{\url{https://github.com/wwaldron/AstrOptical}} and
Astropy~\citep{astropy2013,astropy2018}.
We also include a source table in Table~\ref{tab:baseSrcs}. The table in the paper contains the first
five entries. The full table can be accessed online.

\begin{table*}
    \centering
    \caption{The properties of the \hst{} sources (the first five sources)}
    \label{tab:baseSrcs}
    \begin{tabular}{cccccc}
    \hline
    \#  &            RA &            Dec & F475W & F275W-F475W & F475W-F814W \\
        &   (J2000)     & (J2000)        & (mag) & (mag)       & (mag)       \\
    \hline
    1   &  16:13:20.56 &  -60:46:11.75 &  24.59 $\pm$ 0.09 &  -0.21 $\pm$ 0.23 &  -0.73 $\pm$ 0.28 \\
    2   &  16:13:24.82 &  -60:46:09.76 &  23.35 $\pm$ 0.04 &   1.02 $\pm$ 0.21 &  -0.27 $\pm$ 0.07 \\
    3   &  16:13:24.84 &  -60:46:09.00 &  23.97 $\pm$ 0.06 &   0.55 $\pm$ 0.24 &  -0.03 $\pm$ 0.10 \\
    4   &  16:13:19.77 &  -60:46:11.02 &  24.81 $\pm$ 0.11 &   0.27 $\pm$ 0.44 &   0.16 $\pm$ 0.18 \\
    5   &  16:13:24.70 &  -60:46:08.58 &  23.62 $\pm$ 0.04 &   0.99 $\pm$ 0.25 &  -0.15 $\pm$ 0.08 \\
    \hline
    \end{tabular}
\end{table*}

\subsection{\texorpdfstring{\ion{H}{II}}{HII} regions}
\label{sub:hii}

\begin{figure*}
    \includegraphics[width=1.0\textwidth]{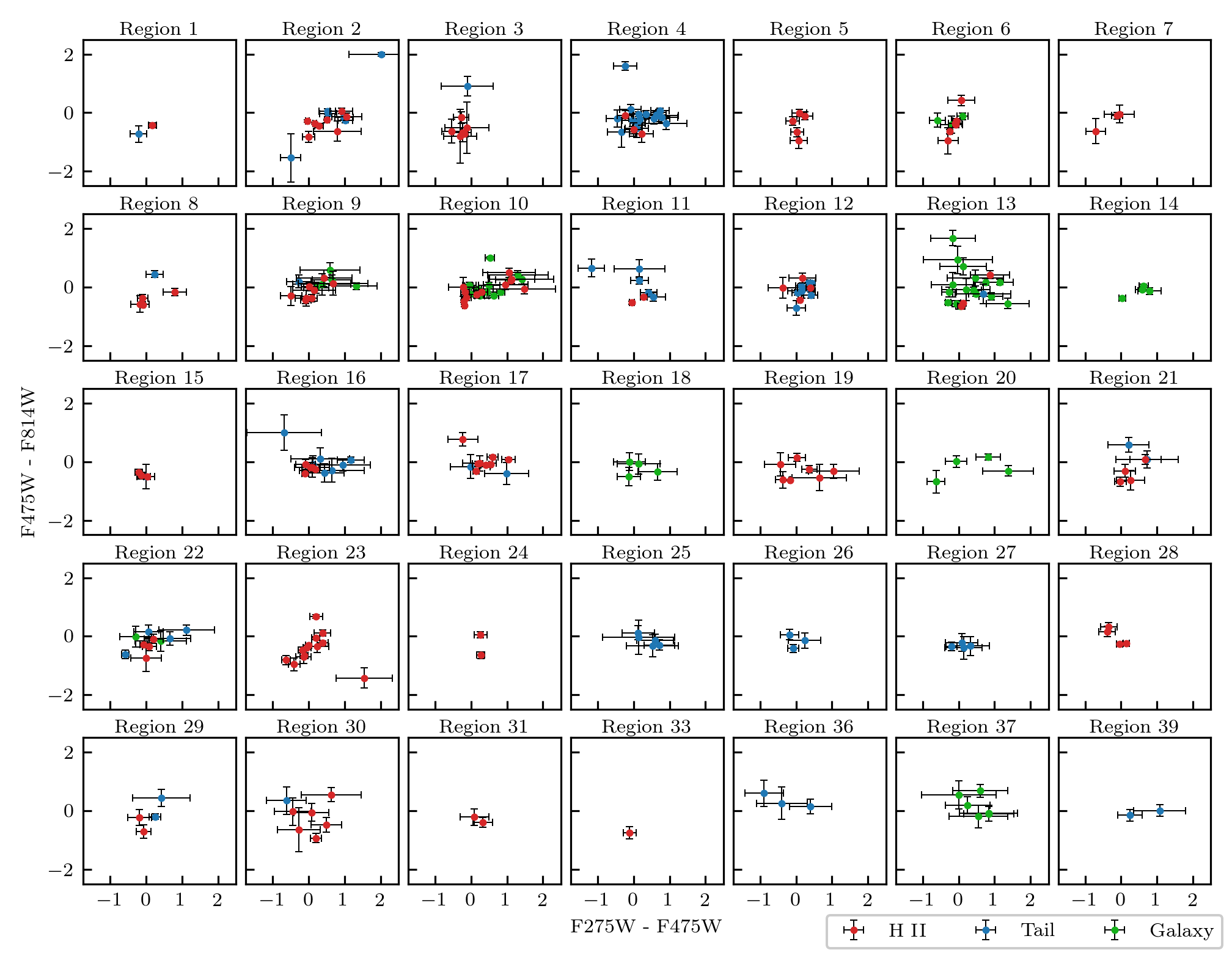}
    \vspace{-0.7cm}
    \caption{
    The colors of individual sources in the clumps defined in Fig.~\ref{fig:wholeWithRegs}.
    Regions selected for CO emission but without any detected \hst{} blue sources are not included here. One can see sources in proximity typically have similar colors.}
    \label{fig:clumpColors}
\end{figure*}

We measured the colors of individual sources found in the clumps defined in Fig.~\ref{fig:wholeWithRegs}.
The results are found in Fig.~\ref{fig:clumpColors}.
The majority of sources shown lie within the \ion{H}{II} regions defined
by \citetalias{sun2007HA} and in this work.
As is shown in the figure, many of the sources in each of the respective clumps
do have similar color within uncertainty regardless of their source category
(i.e., \ion{H}{II}, tail, or galaxy) although there are a few exceptions.
The red tail source in Region~2 is due to the red (likely background) source
in the NW part of the cutout just west of the westernmost \ion{H}{II} region which
was just under the color cutoffs defined in Section~\ref{sub:roi}.
The strong F814W tail source in Region~4 is the source that appears pink
just SW from the center of the cutout.
The high variance of galaxy sources in Region~13 is likely due to the dust gradient
in that clump.

\subsection{SFR Measurement Comparison}
\label{app:sfrComp}

Section~\ref{sub:tailSFH} briefly notes that the \gls{hst} \gls{esfr} is approximately half that of
the \gls{muse} \gls{ew} \gls{sfr}.
Fig.~\ref{app:sfrComp} shows the region-by-region
\gls{sfr} measurement comparison for the \ion{H}{II} regions discussed in Section~\ref{sub:roi} and
Figs.~\ref{fig:compRegions},~\ref{fig:colorComps} and~\ref{fig:museEWvsAV}.
In general, the \gls{muse} \gls{sfr} is higher than the \gls{hst} \gls{esfr}.
The dashed line in this figure shows unity while the dotted line shows the best fit ratio of 0.52.

\begin{figure}
    \centering
    \includegraphics{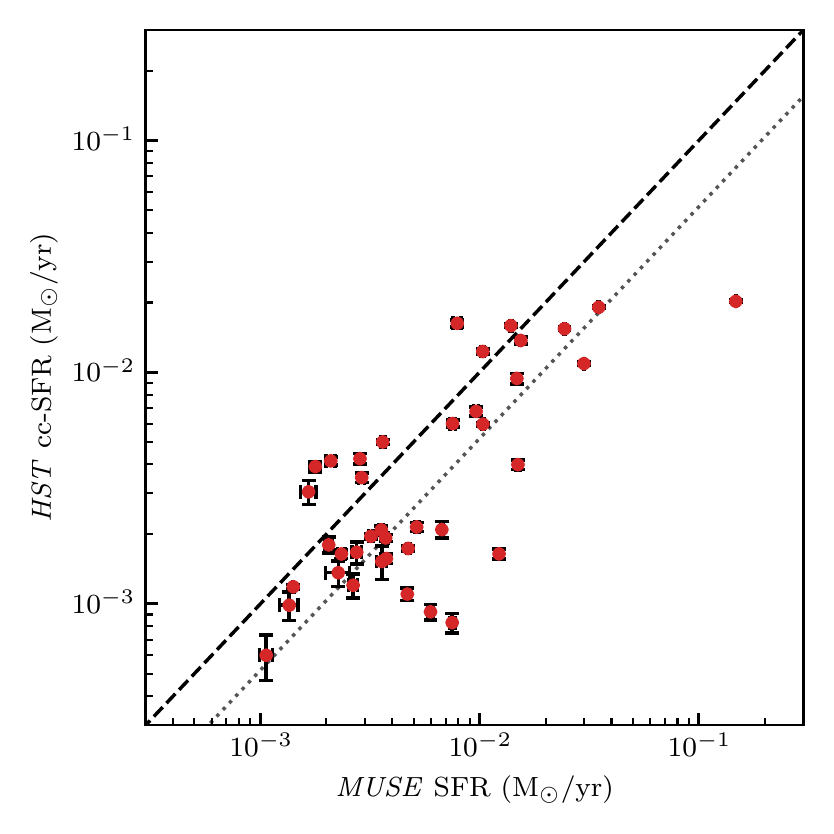}
    \caption{
    The comparison of the \gls{sfr} as measured by the \gls{muse} \gls{ew} and the
    \gls{esfr} as measured by \gls{hst} for the \ion{H}{II} regions discussed in
    Section~\ref{sub:roi} with the exception of four regions where no \gls{hst}
    sources were detected.
    The points represent the median values of the~50,000 Monte-Carlo iterations
    and the error bars represent the median absolute deviation
    (see Section~\ref{sub:tailSFH}).
    The dashed line shows the unity line while the dotted line shows the best fit of the ratio of~0.52.
    The median ratio is~0.61 and the median absolute deviation is~0.24.
    }
    \label{fig:sfrComp}
\end{figure}